\documentclass[fleqn,usenatbib]{mnras}

\usepackage{newtxtext,newtxmath}
\usepackage{longtable}
\usepackage{caption}
\usepackage{afterpage} 
\usepackage{multirow}

\usepackage[T1]{fontenc}

\DeclareRobustCommand{\VAN}[3]{#2}
\let\VANthebibliography\thebibliography
\def\thebibliography{\DeclareRobustCommand{\VAN}[3]{##3}\VANthebibliography}

\usepackage{graphicx}	
\usepackage{amsmath}	
\usepackage{gensymb}    

\usepackage{xcolor}
\usepackage{verbatim}

\definecolor{blue}{HTML}{0766db}

\definecolor{my_color}{HTML}{3a18b1}

\definecolor{new_color}{HTML}{CF0000}

\newcommand\bedit[1]{\textcolor{black}{#1}}
\newcommand\bedittwo[1]{\textcolor{black}{#1}}

\newcommand\notsotiny{\@setfontsize\notsotiny{6}{7}}

\title[Absence of a Correlation between White Dwarf Planetary Accretion and Primordial Stellar Metallicity]{\bedit{Absence of a Correlation between White Dwarf Planetary Accretion and Primordial Stellar Metallicity}}

\author[Jenkins et al.]{Sydney Jenkins,$^{1,2}$\thanks{E-mail: sydneyaj@mit.edu}
Andrew Vanderburg,$^{1}$
Allyson Bieryla,$^{3}$
David W. Latham,$^{3}$
Mariona Badenas-Agusti,$^{1,4}$\newauthor
Perry Berlind,$^{3}$
Simon Blouin,$^{5}$
Lars A. Buchhave,$^{6}$
Michael L. Calkins,$^{3}$
Gilbert A. Esquerdo,$^{3}$
Javier Via\~na$^{1}$
\\
$^{1}$Department of Physics and Kavli Institute for Astrophysics and Space Research, Massachusetts Institute of Technology, Cambridge, MA 02139, USA\\
$^{2}$NSF Graduate Research Fellow\\
$^{3}$Center for Astrophysics $|$ Harvard and Smithsonian, 60 Garden Street, Cambridge, MA 02138, USA\\
$^{4}$Department of Earth, Atmospheric and Planetary Sciences, Massachusetts Institute of Technology,  Cambridge,  MA 02139, USA\\
$^5$Department of Physics and Astronomy, University of Victoria, Victoria, BC V8W 2Y2, Canada\\
$^6$DTU Space,  Technical University of Denmark, Elektrovej 328, DK-2800 Kgs. Lyngby, Denmark
}

\date{Accepted XXX. Received YYY; in original form ZZZ}

\pubyear{2022}

\begin{document}
\label{firstpage}
\pagerange{\pageref{firstpage}--\pageref{lastpage}}
\maketitle

\begin{abstract}
Over a quarter of white dwarfs have photospheric metal pollution, which is evidence \bedit{for recent accretion of exoplanetary material}. While a wide range of mechanisms have been proposed to account for this pollution, there are currently few observational constraints to differentiate between them. 
To investigate the driving mechanism, we observe a sample of polluted and non-polluted white dwarfs in wide binary systems with main-sequence stars. Using the companion stars' metallicities as a proxy for the white dwarfs' primordial metallicities, we compare the metallicities of polluted and non-polluted systems. Because there is a well-known correlation between giant planet occurrence and higher metallicity (with a stronger correlation for close-in and eccentric planets), these metallicity distributions can be used to probe the role of gas giants in white dwarf accretion. We find that the metallicity distributions of polluted and non-polluted systems are consistent with the hypothesis that both samples have the same underlying metallicity distribution. \bedit{However, we note that this result is likely biased by several selection effects. Additionally, we find no significant trend between white dwarf accretion rates and metallicity. These findings suggest that giant planets are not} the dominant cause of white dwarf accretion events in binary systems. 

\end{abstract}



 
\section{Introduction} \label{sec:intro}
Over the past decade, evidence has emerged that 
more than a quarter of white dwarfs have photospheric metal pollution \citep{koe2014}. Because the timescale for these elements to settle into the white dwarfs' interiors is much shorter than the white dwarfs' lifetimes \citep[e.g.,][]{paq1986, koe2009}, the heavy elements must have recently accreted onto the white dwarfs’ surfaces. The elemental abundances of this polluting material 
are generally consistent with the composition of rocky bodies in our own solar system \citep[\bedit{e.g.,}][]{zuc2010}. \bedit{Combined with other evidence, such as the existence of compact dusty debris disks around some polluted white dwarfs \citep[e.g.,][]{zuc1987, gra1990, kil2005} and the atmospheric, spatial and kinematic properties of polluted white dwarfs, which are inconsistent with accretion from the interstellar medium \citep[e.g.,][]{aan1993,far2010}, this strongly suggests that the accreting material originates within the white dwarf's planetary system. We therefore expect it to be common for small rocky bodies to be scattered, tidally disrupted, and subsequently accreted onto the white dwarf photosphere.} This scenario was corroborated by discoveries of  disintegrating rocky bodies transiting white dwarfs \citep[e.g.,][]{van2015, van2020, van2021, gui2021, far2022}.

\begin{figure*}
\centering
\includegraphics[width=\textwidth]{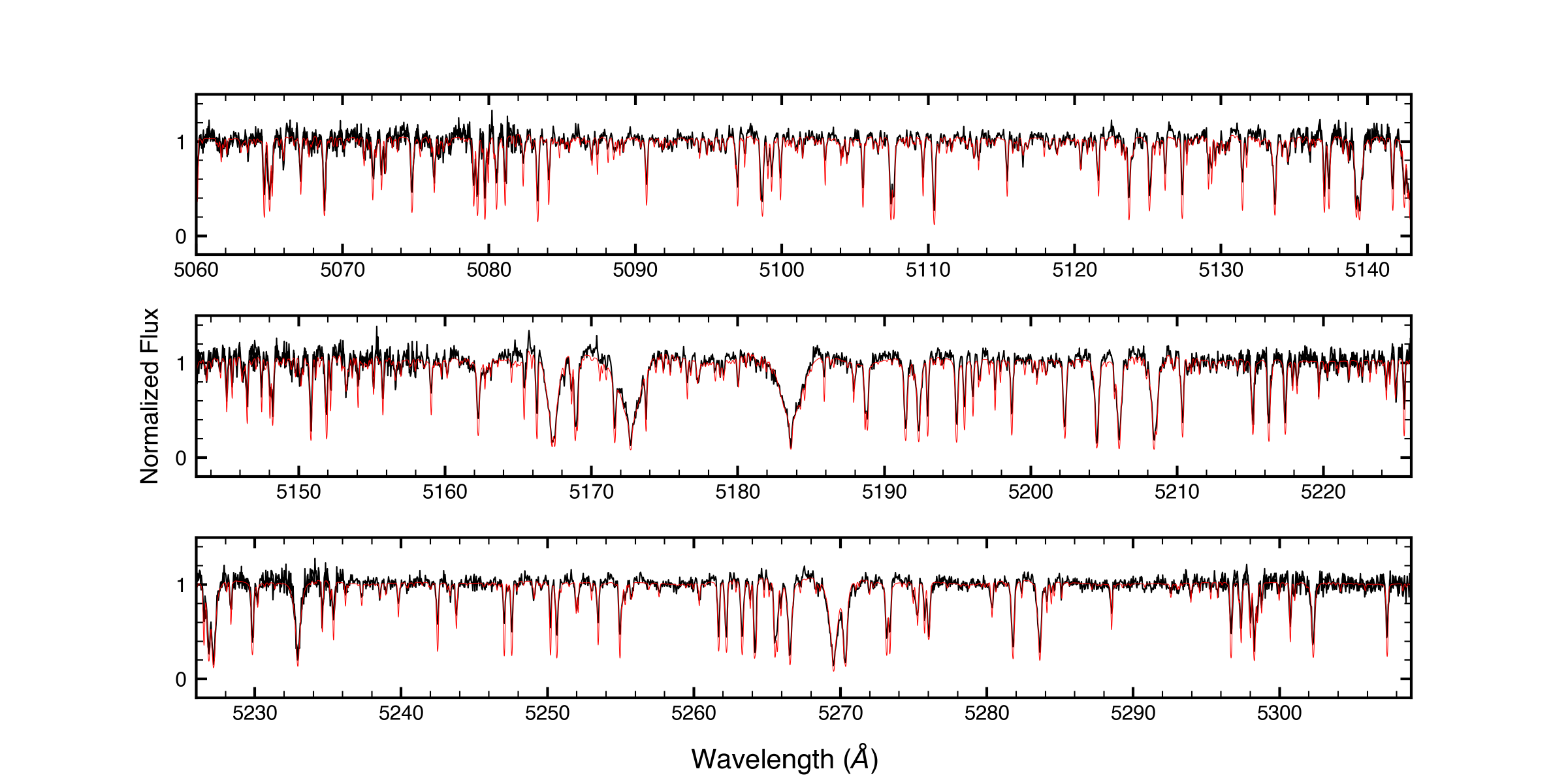}
\caption{Example TRES spectrum of TIC 268127217. We show the wavelength region from 5060 to 5310 \AA, covering the regime used by SPC to calculate stellar metallicities (see \S\ref{sec:SPC}). The TRES data is shown in black, and a corresponding model spectrum is shown in red. We collect TRES spectra of stellar companions to 65 white dwarfs, including 20 polluted white dwarfs and 45 non-polluted white dwarfs. \label{spec}}
\end{figure*}

It is generally believed that rocky bodies are perturbed onto orbits that cross the white dwarfs' Roche radii, where they are tidally disrupted and accreted onto the star. However, the mechanism by which most minor planets are perturbed is still debated, and many different physical processes have been proposed to explain the observed white dwarf accretion rates. Some of these mechanisms do not rely on the presence of a major planet in the system, including sublimation and outgassing \citep{veregg2015}, rotational fission of highly eccentric (e$>$0.99) asteroids \citep{mak2019}, and the Eccentric Kozai-Lidov mechanism in wide binary systems \citep{ste2017}. Additionally, Oort cloud analogues can potentially be disrupted by natal white dwarf kicks triggered by anisotropic mass loss \citep{sto2015} or by galactic tides and stellar flybys \citep{alc1986, par1998, ver2014}.

On the other hand, many other mechanisms do rely on the presence of planets to perturb small bodies towards the white dwarf.
\bedit{A Kuiper-like planetesimal belt with a lower mass planet (< Neptune in size) at a similar semi-major axis can scatter material into the inner planetary system. A sufficient mass can be scattered to explain observed quantities of white dwarf pollution \citep{bon2011, cai2017}. In addition to perturbing objects from Kuiper-belt analogs, planets can also have a similar impact on the orbits of asteroid belt analogs. Rocky bodies in mean motion resonance with a Jupiter-mass planet can be scattered inward and tidally disrupted by the white dwarf \citep{deb2012, ant2019}.} Studies with multiple planets introduce \bedit{further} sources of instability. For instance, \citet{deb2002} showed that planetary systems close to instability may be destabilized when the host star transitions off the main sequence. To explain white dwarf pollution, this mechanism requires a large fraction of planetary systems to be on the cusp of instability. Further work on post-main-sequence systems have investigated the effects of stellar mass loss \citep{voy2013, musver2013,mus2018}, a binary companion \citep{ver2016}, and non-Kozai mutual inclinations \citep{ver2018}. Simulations have also been used to track the stability of two-planet \citep{ver2013, mal2020a} and three-planet \citep{mal2020} systems through the stages of stellar evolution. \citet{mal2020} found that low mass planets (1-100 M$_{\oplus}$) cause instabilities on Gyr timescales, while more massive planets lead to earlier instabilities.

While many pollution \bedit{models} have been proposed, there are currently few observational constraints for distinguishing between \bedit{the different potential physical mechanisms driving accretion.} However, there has been some evidence that white dwarf accretion rates do not strongly decrease over several Gyrs \citep{blo2022}. Giant planets, which tend to pollute earlier in a white dwarf's life, may therefore be unable to explain the observed mass accretion rates. To further probe the importance of massive planets in driving pollution, we measure the metallicity of white dwarf wide visual binary systems. \bedit{This approach relies on two well-established relations: the planet-metallicity correlation and the chemical homogeneity of stars in binary systems.}

For main sequence stars, there is a known correlation between the host star metallicity and the presence of giant exoplanets \citep{fis2005}. This correlation is not as strong for lower mass planets \citep{buc2012}. If giant planets are required to perturb small rocky bodies inward to pollute white dwarfs, there should therefore be a correlation between white dwarfs' primordial metallicities and the presence of heavy element pollution. If giant planets are not required, 
the primordial metallicity \bedit{distributions of polluted and non-polluted white dwarfs should be indistinguishable.} We note that hot and eccentric Jupiters may tend to form in higher metallicity environments than cool Jupiters \citep{buc2018}, possibly because such systems tend to host multiple giant planets, leading to more planet-planet interactions. However, as discussed in \S\ref{sec:prediction_comparison}, we still expect there to be a metallicity offset between systems with no Jupiters and those with cool Jupiters. 
\begin{figure}
\centering
\includegraphics[width=\columnwidth]{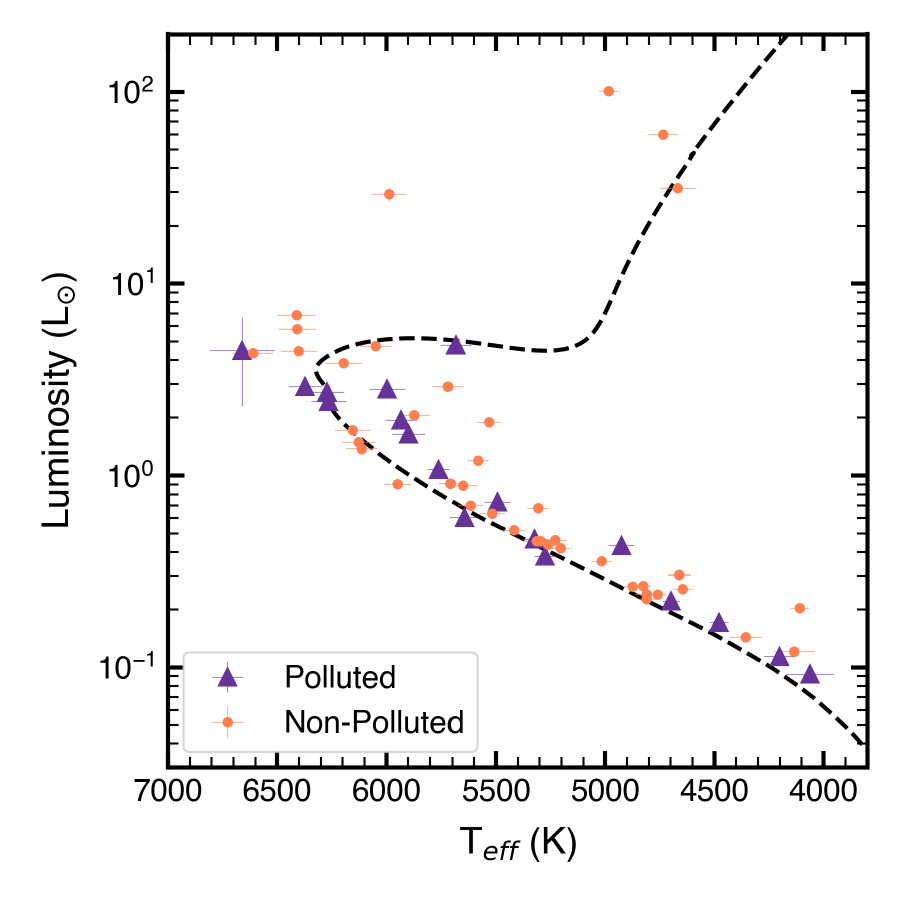}
\caption{Hertzsprung-Russell diagram of white dwarf companion stars. Luminosities and effective temperatures are calculated using the {\tt Isochrones} package, as described in \S\ref{sec:SPC}. A MIST isochrone is shown in black.  \label{HR}}
\end{figure}

\begin{figure*}
\centering
\includegraphics[width=.95\textwidth]{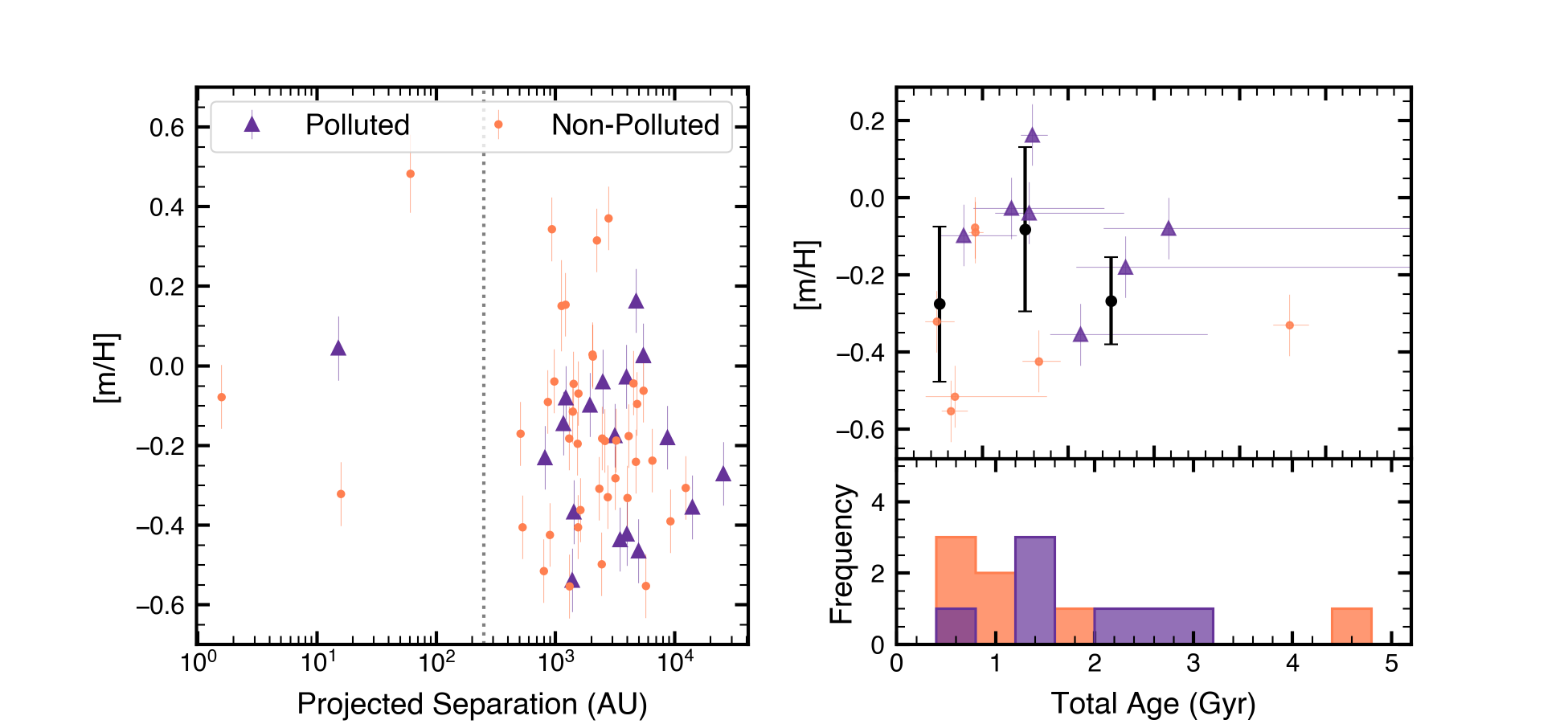}
\caption{Characteristics of binary systems observed in this study. Left: Distribution of metallicities and \bedit{projected} separations between stars in binary systems\bedit{, with a vertical dashed line corresponding to 250 au.} Close-in binaries \bedit{are known to} display a correlation between separations and metallicity \bedit{\citep{elb2019}}. However, when we remove systems with separations $< 250$ \bedit{au} from our sample, our results do not significantly change, indicating that the separation-metallicity correlation does not bias our results. Right: Distribution of white dwarf ages, as described in \S\ref{sec:relations}. We observe no significant trend in metallicity as a function of age, indicating that the systems' ages also do not bias our results. \label{characteristics}}
\end{figure*}

It is generally impossible to measure the primordial metallicity of a white dwarf, as the white dwarf’s strong gravity causes primordial metals to sink to the core, leaving envelopes composed of either pure hydrogen or helium. However, white dwarfs in visual binaries with main sequence stars provide a way to infer the white dwarf primordial metallicities, as the two stars are coeval and presumably formed from gas of a similar composition \citep{haw2020}. The metallicity of companions to both polluted and non-polluted white dwarfs can therefore be used to probe the strength of any white dwarf pollution/metallicity correlation and place preliminary constraints on how planetesimals are perturbed towards their host stars \bedit{\citep{bon2021}}. 

In this work, we probe the importance of large planets in driving white dwarf pollution by measuring the metallicities of binary systems containing polluted and non-polluted white \bedit{dwarfs. 
Our} paper is organized as follows: we describe our sample selection, observations, and metallicity measurements in \S\ref{sec:method}. We then compare the metallicity distributions \bedit{and inferred accretion rates} of polluted and non-polluted systems in \S\ref{sec:Results}\bedit{, and describe several experimental caveats in \S\ref{sec:caveats}. We propose strategies for addressing these caveats, in addition to potential future estimates of Jupiter frequencies, in \S\ref{sec:future}. We then discuss constraints on the dominant white dwarf pollution mechanisms in \S\ref{sec:discussion}, before concluding in \S\ref{sec:conclusion}.} 

\section{Method} \label{sec:method}
\subsection{Sample Selection}
\label{sec:data}
We draw our sample \bedit{of wide binaries with both a main sequence star and white dwarf companion} from two sets of targets. We first use observations collected in a 2017B Tillinghast Reflector Echelle Spectrograph \citep[TRES,][]{gaborthesis} observing program. \bedit{The targets for this program} were identified before the {\it Gaia} DR2 catalog \citep{gaia2016, gaia2018} became available. \bedit{We therefore selected them from catalogs of known white dwarf/main sequence visual binaries  \citep{hol2013, zuc2014}.} 
To increase our sample of metal polluted white dwarfs, we searched $\sim$350 known polluted white dwarfs \citep{duf2007,koe2015,hol2017} for common proper motion companions. As most of these white dwarfs were identified by the Sloan Digital Sky Survey (SDSS), we used the proper motions measured by the SDSS pipeline using the offset between USNO-B positions \bedit{\citep{mon2003}} and SDSS positions. We then queried proper motions for all stars in the UCAC-4 catalog \bedit{\citep{zac2013}} within 10 arcminutes of the white dwarfs and searched for stars with similar proper motions\bedit{. Following the release of the {\it Gaia} \bedit{Early Data Release 3 (EDR3)} catalogue \citep{pru2016, bro2020}, we used updated proper motions to verify that our targets are binary systems. Stars with right ascension and declination proper motion components within $\sim$3 mas yr$^{-1}$ are considered true binaries.} 
Our final 2017 sample includes 6 polluted systems and 34 non-polluted systems. 

Our second set of targets \bedit{is} selected using \bedit{{\it Gaia} EDR3} data to identify white dwarfs in binaries. We \bedit{cross-match} polluted white dwarfs in the Montreal White Dwarf \bedit{Database \citep[MWDD,][]{duf2016}} and in LAMOST \citep{luo2015, kon2021, mar2023} with a catalog of \bedit{{\it Gaia} EDR3} binaries \citep{elb2021} using the targets' \bedit{{\it Gaia} EDR3} IDs. We \bedit{restrict}
our sample to targets with a declination  greater than $-$30$^\circ$, apparent \bedit{{\it G}}-band magnitude less than 14, and \bedit{corresponding absolute magnitude} less than 8. We also \bedit{require} that all targets have an effective temperature greater than 4000 K, as our metallicity estimates (described in \S\ref{sec:SPC}) are only valid in this range. 
We \bedit{identify} a total of 14 polluted targets, in addition to 11 non-polluted targets. \bedit{However, we note that \bedit{MWDD} is not complete for polluted white dwarfs, and that our sample may therefore be biased by unknown selection effects. }

Our final sample, \bedit{including both the targets observed in the 2017B TRES program and the new targets identified in MWDD}, includes 20 polluted and 45 non-polluted systems. 
An example spectrum \bedit{of a white dwarf's stellar companion is shown in Figure~\ref{spec}.} Figure~\ref{HR} shows an HR diagram of the binary companions to the white dwarf stars, while Figure~\ref{characteristics} shows the separation, age, and metallicity distributions of our target systems. 

\subsection{TRES Observations} 
We use observations from the Smithsonian Astrophysical Observatory's TRES echelle spectrograph on the 1.5-meter Tillinghast telescope. Our targets were observed between December 2009 and February 2023, with a median signal-to-noise ratio \bedit{(SNR) of 83.} Observations prior to 2017 were part of a different observing program. An example TRES spectrum of TIC 268127217 is shown in Figure~\ref{spec}, covering the wavelength regime from 5060 to 5310 \AA. 

We observe six stars \bedit{with} unreliable metallicities. Three are in double-line spectroscopic binaries (TICs 435872531, 8243025, and 85321782) and three are rapid rotators (TICs 67401178, 302187609, and 456863710). Two of these (TICs 85321782 and 302187609) are companions to polluted white dwarfs. We do not include these targets in the  \bedit{following analysis, restricting our sample to the remaining 18 polluted and 41 non-polluted systems.} 

\subsection{Stellar Parameter Classification}
\label{sec:SPC}
We obtain metallicities using Stellar Parameter Classification (SPC), as documented by \citet{buc2012}. SPC cross-correlates an observed spectrum against a grid of model template spectra. The model spectra, calculated using the \citet{kur1992} grid of model atmospheres, cover the wavelength region between 5050 and 5360 \AA\, and correspond to varying effective temperature $T_{\textrm{eff}}$, surface gravity \bedit{log($g$)}, metallicity [m/H], and projected equatorial rotational velocity $V_{\textrm{rot}}$. The synthetic spectra span the following ranges: 3500 K $< T_\textrm{eff} <$  9750 K, 0.0 $<$ log($g$) $<$ 5.0, $-$2.5 $<$ [m/H] $<$ +0.5, and 0 km s$^{-1}$ $< V_{\textrm{rot}} <$ 200 km s$^{-1}$. The synthetic spectra library has a spacing of 250 K in effective temperature, 0.5 in log($g$), 0.5 dex in [m/H], and a spacing between 1 to 20 km s$^{-1}$ in rotational velocity. The final library includes a total of 51,359 spectra. The normalized cross-correlation function (CCF) peak, a measure of how well a synthetic template fits the observed spectrum, is then used to determine the value of each stellar parameter and to estimate internal uncertainties. \bedit{See \citet{fur2018} for a comparison of stellar parameters produced by SPC and those produced by three other data analysis pipelines: {\tt Kea} \citep{end2016}, {\tt Newspec} \citep{eve2013}, and {\tt SpecMatch} \citep{pet2017}.}

If additional information is known about one of the stellar \bedit{parameters}, a prior can be included in SPC. To improve our measurements, we calculate a prior for \bedit{log($g$)}, a value that is difficult to determine spectroscopically \citep[e.g.,][]{tor2012}. We use {\tt Isochrones}, a Python package for stellar evolution modelling \citep{mor2015}. For each target, we provide the SPC metallicity, effective temperature, and photometry from {\it Gaia}, 2MASS \citep{skr2006}, and WISE \citep{wri2010}. We also include parallaxes from {\it Gaia} \bedit{EDR3} when possible, and use values from Hipparcos when {\it Gaia} \bedit{EDR3} measurements are unavailable. We account for dust reddening using the Bayestar dust map developed by \citet{gre2019}. We then use {\tt Isochrones} to estimate \bedit{mass,} radius, and \bedit{log($g$)}. 

\begin{figure}
\centering
\includegraphics[width=1.1\columnwidth]{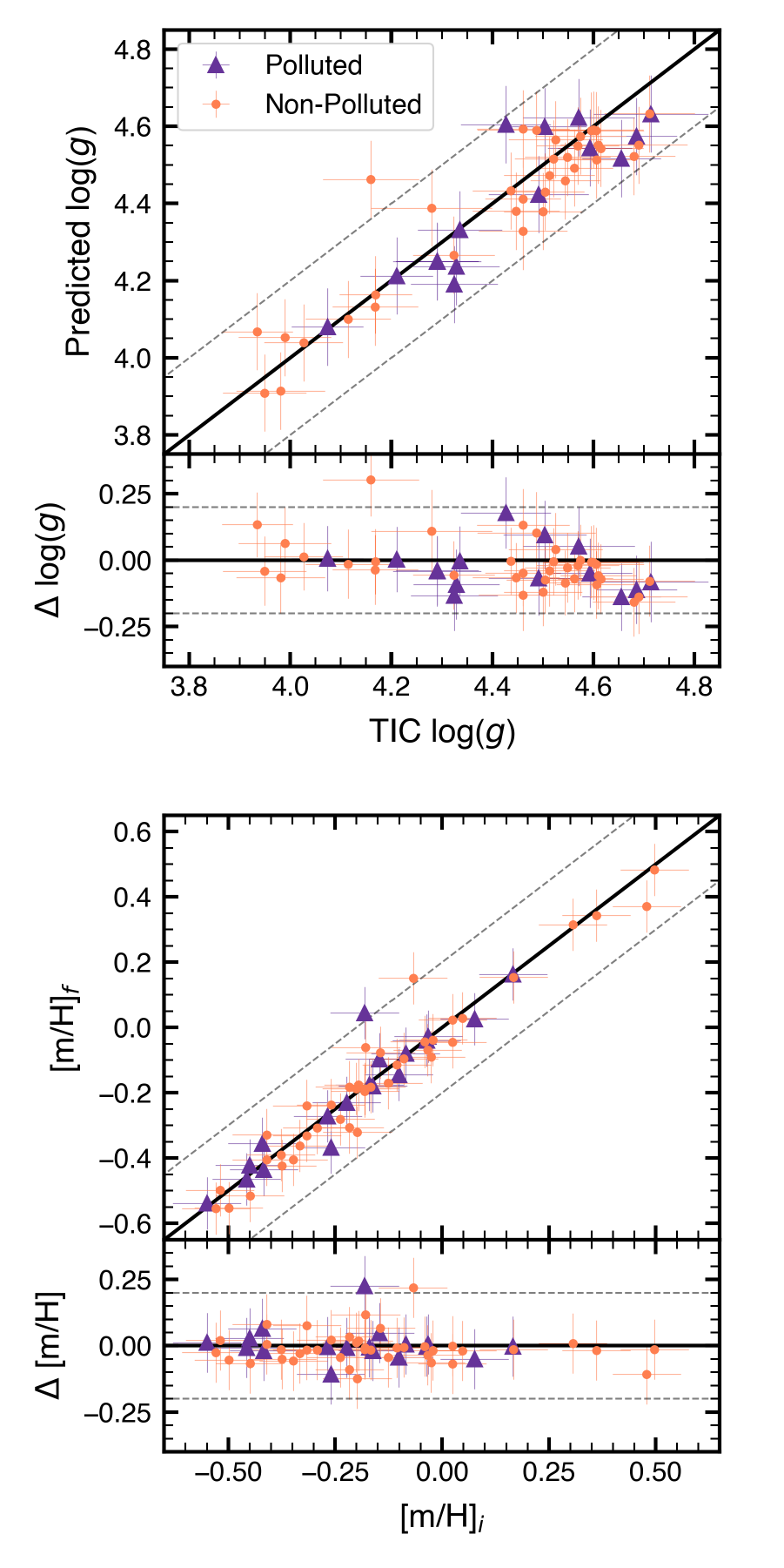}
\caption{Top: Comparison of log($g$) values measured in this paper and available in TIC. Our log($g$) values (on average $0.02\pm0.01$ lower than the TIC values) are used to improve the SPC calculations of metallicity (see \S\ref{sec:SPC}). Bottom: Comparison of metallicity values before ($[$m/H$]_i$) and after ($[$m/H$]_f$) applying log($g$) priors to SPC. The values are consistent with each other, with a mean difference of $0.00\pm0.01$ dex. We use the final SPC metallicity values to compare the metallicity distributions of polluted and non-polluted white dwarf systems. \label{logg}}
\end{figure}

We compare our log($g$) results with the values in the TESS Input Catalog \citep[TIC,][]{sta2019}, as shown in the upper panel of Figure~\ref{logg}. We find that our log($g$) values are on average $0.02\pm0.01$ lower than the TIC values. Using our updated log($g$) values to establish priors, we re-run SPC. In order to determine whether this prior has a significant effect on the estimated metallicities, we compare our initial and final SPC results in the bottom panel of Figure~\ref{logg}. The mean metallicity difference \bedit{$\mu$([m/H]$_f -$[m/H]$_i$)} is 0.00$\pm$0.01 dex, indicating that our initial and final metallicity measurements are consistent with each other. \bedit{However, we note that the corresponding standard deviation $\sigma$([m/H]$_f$-[m/H]$_i$) is 0.063$\pm$0.006.} Our final metallicity measurements are given in Table~\ref{tab:SPClist}. 

\begin{table*}
\centering
\caption{SPC results for white dwarf stellar companions. We include the companions' TIC IDs, positions, and mean signal\bedit{-}to\bedit{-}noise ratios (SNRs) per resolution element at the middle of the Mg b region, in addition to the SPC-calculated metallicity, effective temperature, surface gravity, and projected rotational velocity (see \S\ref{sec:SPC}). Unless otherwise indicated, the corresponding uncertainties are 0.08 dex, 50 K, 0.1 cgs, and 0.5 km s$^{\bedit{-}1}$, respectively. } \label{tab:SPClist}
\centering
\begin{tabular}{cccccccc}
\hline
\hline
 Companion ID & RA & Dec& SNR & [m/H] & T$_{\textrm{eff}}$ & log(g) & $V_{\textrm{rot}}$ \\
 & (deg) & (deg) &   & (dex) & (K) & (cgs) & (km s$^{\bedit{-}1}$) \\
\hline
376455973 & 8.182083 & 8.457583 & 20.0 & \bedit{$-$}0.37 & 4220 & 4.6 & 2.6 \\ 
 191145234 & 8.255833 & 44.730000 & 25.4 & \bedit{$-$}0.19 & 5240 & 4.5 & 1.0 \\ 
 268127217 & 11.337500 & 14.363056 & 36.2 & \bedit{$-$}0.46 & 5250 & 4.6 & 0.5 \\ 
 336892483 & 18.033333 & 4.916167 & 56.4 & \bedit{$-$}0.10 & 6260 & 4.2 & 7.8 \\ 
 35703955 & 23.990417 & \bedit{$-$}0.448444 & 43.2 & \bedit{$-$}0.54 & 5620 & 4.5 & 1.2 \\ 
 270371546 & 32.417083 & 5.949222 & 34.2 & \bedit{$-$}0.27 & 5800 & 4.2 & 3.8 \\ 
 129904995 & 36.170000 & 40.139722 & 27.8 & \bedit{$-$}0.07 & 4770 & 4.6 & 1.8 \\ 
 251083146 & 37.515000 & \bedit{$-$}0.377500 & 26.2 & 0.34 & 5330 & 4.4 & 3.1 \\ 
 422890692 & 37.862083 & 2.757861 & 32.8 & \bedit{$-$}0.14 & 5470 & 4.4 & 2.7 \\ 
 439869953 & 43.380417 & \bedit{$-$}0.568889 & 57.5 & \bedit{$-$}0.18 & 5330 & 4.6 & 0.9 \\ 
 381347690 & 46.767083 & 15.675556 & 27.5 & \bedit{$-$}0.55 & 4380 & 4.6 & 2.1 \\ 
 279049424 & 49.593333 & \bedit{$-$}0.930278 & 120.7 & \bedit{$-$}0.28 & 4640 & 2.5 & 3.7 \\ 
 435915513 & 65.219583 & 13.864444 & 107.0 & 0.48$\pm$0.10 & 6210$\pm$100 & 4.1$\pm$0.2 & 23.0$\pm$0.7 \\ 
 429102184 & 66.156250 & 33.959722 & 156.5 & \bedit{$-$}0.32 & 6410 & 3.9 & 35.8 \\ 
 445709402 & 68.316667 & 55.462139 & 96.2 & \bedit{$-$}0.31 & 6130 & 4.0 & 8.1 \\ 
 125838647 & 69.200833 & 27.132222 & 34.5 & 0.32 & 4920 & 4.6 & 7.0 \\ 
 91690128 & 88.494583 & 12.419444 & 93.6 & \bedit{$-$}0.10 & 5850 & 4.2 & 3.5 \\ 
 53088536 & 89.007500 & 56.041389 & 36.0 & \bedit{$-$}0.36 & 5190 & 4.5 & 1.5 \\ 
 171848237 & 96.966250 & 35.003333 & 79.0 & 0.02 & 5620 & 4.4 & 1.6 \\ 
 172703279 & 101.115000 & \bedit{$-$}28.545278 & 37.3 & \bedit{$-$}0.17 & 4880 & 4.6 & 1.3 \\ 
 417756881 & 106.012083 & 71.194722 & 55.6 & \bedit{$-$}0.50 & 5580 & 4.4 & 1.4 \\ 
 280310048 & 114.825417 & 5.225000 & 249.7 & 0.04 & 6700 & 4.3 & 6.6 \\ 
 293414051 & 131.937083 & 12.887361 & 76.0 & \bedit{$-$}0.08 & 5920 & 4.2 & 3.2 \\ 
 241194061 & 153.287083 & 27.419583 & 83.9 & \bedit{$-$}0.04 & 5500 & 4.1 & 3.6 \\ 
 1732730 & 156.212917 & \bedit{$-$}0.401861 & 92.1 & 0.16 & 6210 & 4.2 & 6.7 \\ 
 374210257 & 158.484167 & 2.966667 & 88.6 & \bedit{$-$}0.24 & 6060 & 3.9 & 9.0 \\ 
 17301096 & 160.494583 & 41.187389 & 74.8 & \bedit{$-$}0.36 & 6350 & 4.2 & 7.7 \\ 
 47350711 & 162.496667 & \bedit{$-$}0.791667 & 73.2 & 0.37 & 5750 & 4.1 & 4.0 \\ 
 20324932 & 167.375417 & \bedit{$-$}25.990833 & 77.1 & \bedit{$-$}0.33 & 5430 & 4.5 & 0.7 \\ 
 239209564 & 170.360417 & 14.292722 & 76.2 & \bedit{$-$}0.42 & 5660 & 4.3 & 1.5 \\ 
 138545439 & 174.149167 & 61.666389 & 34.7 & \bedit{$-$}0.04 & 5060 & 4.5 & 2.4 \\ 
 176879529 & 183.094583 & \bedit{$-$}6.359722 & 46.5 & \bedit{$-$}0.39 & 4720 & 4.6 & 1.2 \\ 
 347389665 & 196.808750 & 22.456944 & 58.5 & \bedit{$-$}0.04 & 5520 & 4.5 & 1.3 \\ 
 458455132 & 209.272917 & 33.806944 & 77.3 & \bedit{$-$}0.06 & 6140 & 4.3 & 5.2 \\ 
 445863106 & 215.343333 & 57.067639 & 36.0 & \bedit{$-$}0.18 & 5360 & 4.5 & 1.7 \\ 
 284628177 & 216.904583 & 53.796944 & 28.6 & \bedit{$-$}0.44 & 4230 & 4.6 & 1.9 \\ 
 420807841 & 220.698333 & 6.590556 & 98.4 & 0.03 & 5800 & 3.9 & 6.1 \\ 
 229874145 & 224.473333 & 29.875833 & 37.0 & \bedit{$-$}0.11 & 5290 & 4.5 & 0.7 \\ 
 160424250 & 235.546250 & 72.788611 & 21.8 & 0.15$\pm$0.12 & 5430$\pm$120 & 4.6$\pm$0.2 & 11.9$\pm$0.9 \\ 
 149692037 & 245.529583 & 12.214444 & 77.9 & \bedit{$-$}0.19 & 6060 & 4.4 & 5.7 \\ 
 207468713 & 246.072500 & 55.331972 & 122.7 & \bedit{$-$}0.23 & 5940 & 4.1 & 4.0 \\ 
 318811655 & 246.547500 & 11.407500 & 129.8 & \bedit{$-$}0.31 & 4770 & 2.4 & 4.2 \\ 
 5876391 & 246.572917 & 2.179167 & 32.8 & \bedit{$-$}0.40 & 4760 & 4.5 & 1.5 \\ 
 162690651 & 246.882083 & 49.002750 & 55.6 & \bedit{$-$}0.18 & 4460 & 4.6 & 2.3 \\ 
 21860383 & 257.205000 & 33.215278 & 63.5 & \bedit{$-$}0.18 & 6090 & 4.4 & 4.8 \\ 
 219857965 & 257.467500 & 68.333889 & 26.9 & \bedit{$-$}0.20 & 4280 & 4.6 & 2.4 \\ 
 377626776 & 264.741250 & 13.329167 & 196.2 & \bedit{$-$}0.52 & 6150 & 2.9 & 96.5 \\ 
 277255071 & 268.235833 & 9.803889 & 53.7 & 0.15 & 5070 & 4.6 & 1.1 \\ 
 23395123 & 275.299167 & 32.877861 & 130.2 & \bedit{$-$}0.24 & 6490 & 4.1 & 11.4 \\ 
 229770891 & 282.200417 & 68.876667 & 49.4 & \bedit{$-$}0.33 & 5930$\pm$60 & 4.5$\pm$0.1 & 2.7 \\ 
 76956031 & 318.234167 & 30.226944 & 242.4 & \bedit{$-$}0.08 & 5060 & 2.6 & 4.8 \\ 
 352817358 & 323.049583 & 0.221667 & 37.2 & \bedit{$-$}0.55 & 4810 & 4.5 & 9.9 \\ 
 305267214 & 324.083750 & 11.625389 & 26.1 & \bedit{$-$}0.03 & 4790 & 4.6 & 2.7 \\ 
 259014661 & 334.893333 & 21.367222 & 28.3 & \bedit{$-$}0.18 & 4880 & 4.6 & 1.7 \\ 
 36726370 & 343.958333 & \bedit{$-$}7.822500 & 88.5 & \bedit{$-$}0.40 & 5610 & 4.5 & 1.2 \\ 
 427863040 & 345.282083 & 40.938889 & 21.1 & \bedit{$-$}0.42 & 4200 & 4.4 & 5.6 \\ 
 428986692 & 345.647917 & 76.503889 & 49.7 & \bedit{$-$}0.09 & 5700 & 4.3 & 18.1 \\ 
 91102009 & 349.902500 & 27.490444 & 133.8 & 0.03 & 6330 & 4.1 & 11.9 \\ 
 9726029 & 358.369583 & \bedit{$-$}8.071944 & 32.1 & \bedit{$-$}0.04 & 5310 & 4.5 & 1.7 \\ 
\hline
\end{tabular}
\end{table*}

\section{Results} 
\label{sec:Results}
\subsection{Metallicity Distributions} 
\label{metdistr}
\begin{figure*}
\centering
\includegraphics[width=.98\textwidth]{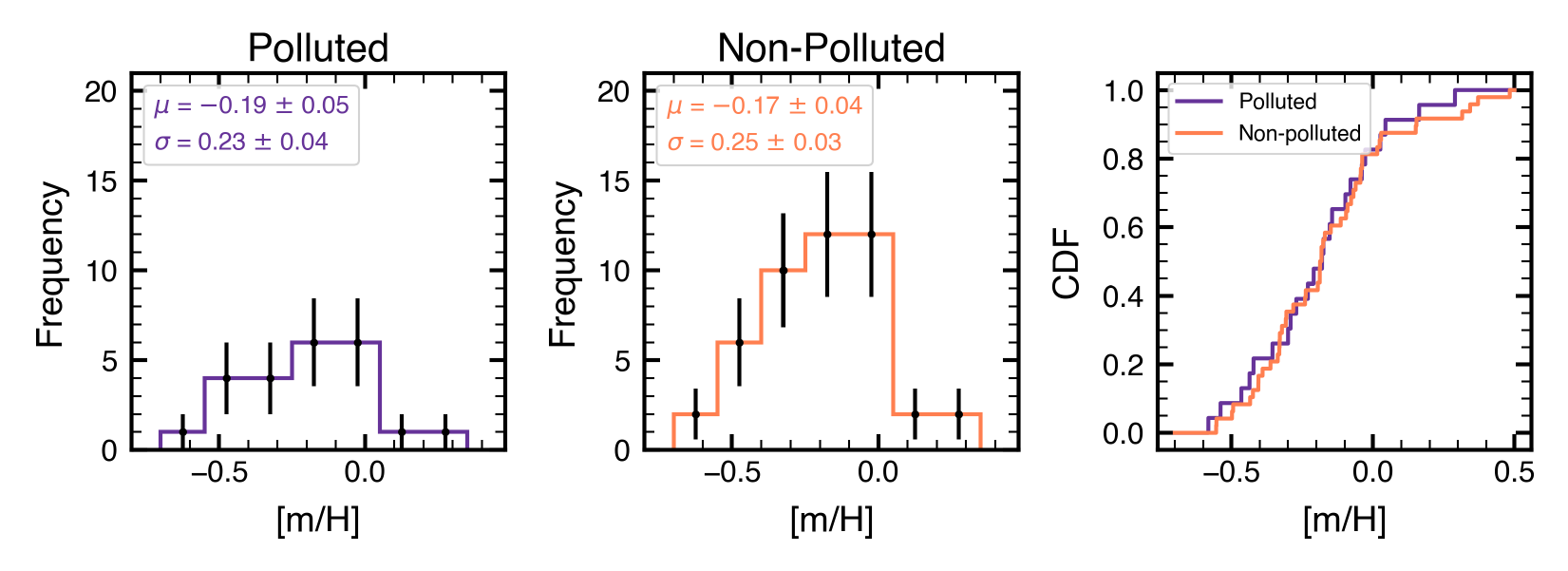}
\caption{Metallicity distributions of white dwarf systems. Left: Metallicity distribution of polluted systems, with error bars shown in black. Middle: Metallicity distribution of non-polluted systems, with error bars shown in black. Right: CDFs of non-polluted and polluted systems, shown in orange and purple, respectively. We find that the mean metallicity for non-polluted systems is consistent with the mean metallicity for polluted systems, indicating that there is not a significant metallicity offset between the two populations and that giant planets are therefore not dominant drivers of white dwarf pollution.\label{met_distr}}
\end{figure*}

We fit a Gaussian distribution to both the polluted and non-polluted metallicity distributions using a Markov Chain Monte Carlo (MCMC) sampler. We calculate a mean and standard deviation of $-0.20\pm0.05$ and $0.22\pm0.04$ dex for the polluted systems, and $-0.16\pm0.04$ and $0.26\pm0.03$ dex for the non-polluted systems. The mean values are therefore consistent within uncertainties. 
The metallicity distributions and corresponding cumulative density functions (CDFs) are shown in Figure~\ref{met_distr}.

We also conduct a two-sample Kolmogorov-Smirnov (KS) test to determine the likelihood that the polluted and non-polluted metallicity distributions are drawn from the same probability distribution. We calculate a KS statistic of 0.11, with a corresponding p-value of 0.99. We therefore find no evidence that the two samples are drawn from different metallicity distributions. This implies that massive planets are not strongly required to pollute white \bedit{dwarfs. However, we note several caveats to this conclusion in the following sections (\S\ref{sec:accretion} and \S\ref{sec:caveats}).}

\subsection{Impact of Giant Planets on Accretion Rates}
\label{sec:accretion}
\bedit{Our result in \S\ref{metdistr} is limited by the sensitivity of our polluted and non-polluted samples. 
The level of pollution is known to depend on the white dwarf's effective temperature \citep{hol2017,cou2019,blo2022} and atmospheric composition \citep[e.g.,][]{dre1996, goo2005, ven2006, bar2014}. Notably, it is easier to detect pollution signatures in a Helium-dominated  white dwarf atmosphere than in a Hydrogen-dominated white dwarf atmosphere. To compare the sensitivities of our two samples, we use H/He ratios measured from spectra, available for 46 targets. We use the visual spectral classification for the remaining 13 systems.  While our polluted sample contains 12 (out of 18) He-dominated white dwarfs, our non-polluted sample contains just 7 (out of 41). It is therefore possible that some of our "non-polluted" white dwarfs are in fact polluted, but have not been observed with sufficient sensitivity to detect pollution signatures. }

\bedit{To work around this difference in sensitivities, we investigate whether there is a correlation between the primordial stellar metallicity and the accretion rate of planetary material onto the white dwarf (rather than the presence of any pollution at all). We estimate inferred accretion rates based on each white dwarf's spectral class, mass, and temperature.}
We use atmospheric parameters and Ca abundances reported in the literature \citep{zuc2010,cou2019,blo2022,farihi2022,car2023}, in addition to diffusion timescales and convection zone masses from \cite{koester2020}. \bedit{For three \bedittwo{polluted} targets (TICs 284628173, 630273894, and 841989747), we estimate Ca abundances ourselves using the available spectra. \bedittwo{These Ca abundances are obtained using the model atmospheres described in \citet{blo2018a, blo2018b}. The effective temperatures and surface gravities are fixed to the values given in Table~\ref{tab:targetlist}, and the Ca/H(e) abundance are adjusted to fit optical spectra in the Ca II H and K region. For stars where no metal lines are detected, the upper limit is determined visually by comparing models with different Ca/H(e) abundances to the available optical spectra.} As in \citet{far2012}, we assume that Ca accounts for 1.6\% of the accreted material when converting Ca accretion rates to total accretion rates. \bedittwo{This is consistent with an Earth-like bulk composition \citep{all1995}, and is therefore not representative of all pollution compositions.} For non-polluted white dwarfs with available spectra, we calculate an upper limit on the accretion rate. We provide our total inferred accretion rates and upper limits in Table~\ref{tab:accretion}, in addition to Ca abundances and sinking timescales. }

\begin{figure}
\centering
\includegraphics[width=\columnwidth]{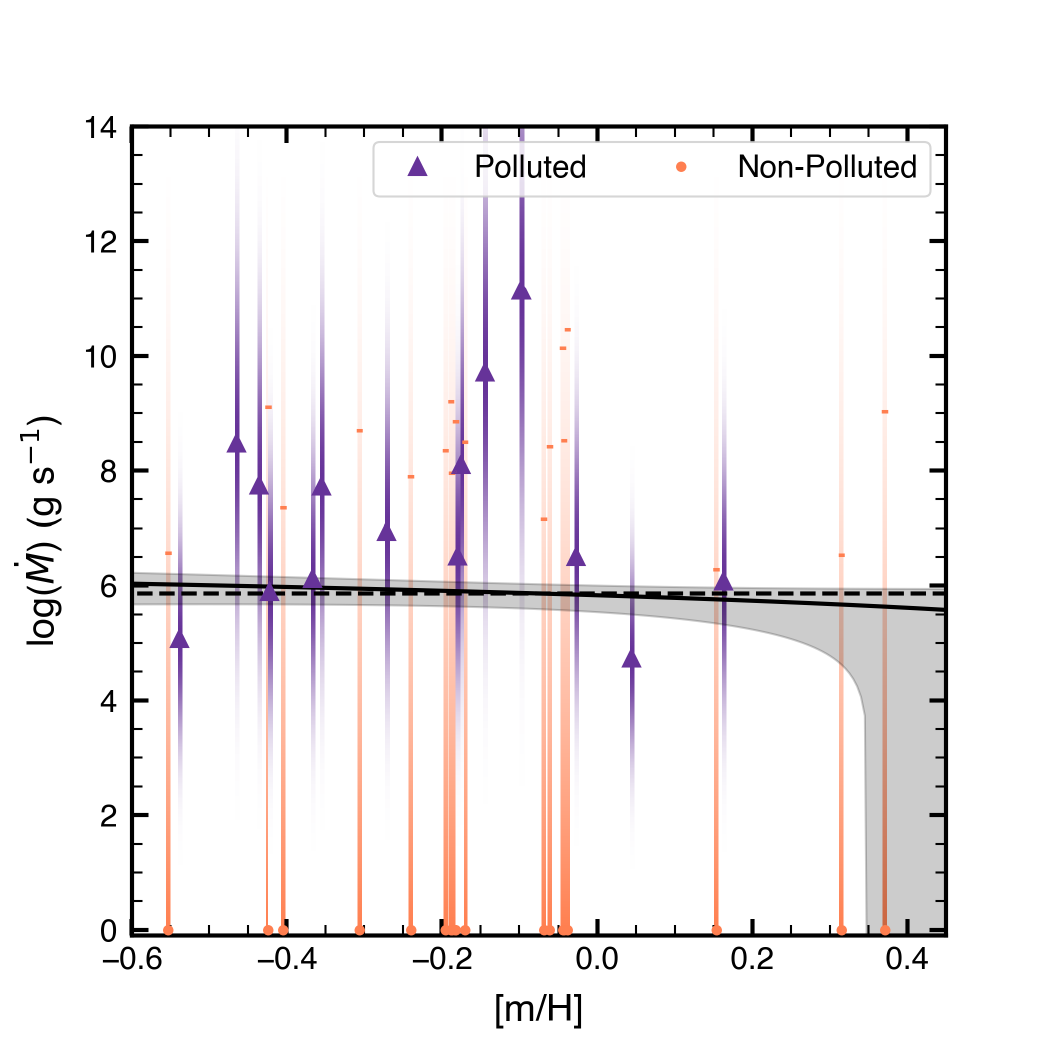}
\caption{\bedit{Total inferred accretion rates as a function of metallicity. We estimate accretion rates from Ca abundances and estimate upper limits for non-polluted targets (see \S\ref{sec:accretion}). We show uncertainties with color proportional to the probability density function. Orange rectangles correspond to the 3$\sigma$ upper limit for non-polluted systems. We find the best-fit line, denoted by the solid black line, with an MCMC sampler. 1$\sigma$ uncertainties for the fit are shown in gray. The slope is consistent with a flat line, such as the horizontal dashed line. We therefore do not find evidence for a relationship between inferred accretion rate and metallicity. Accretion rates are provided in Table~\ref{tab:accretion}. } \label{accretion2}}
\end{figure}

\bedit{To search for a dependence between total accretion rates and primordial metallicity, we use an MCMC sampler with a three-parameter likelihood. We estimate the slope and intercept of accretion rates as a function of metallicity, in addition to a jitter term used to correct the uncertainties of polluted targets. For polluted sources, we fit the estimated accretion rates $R_{\mathrm{P}}$ and corresponding uncertainties $\sigma_{\mathrm{P}}$ as a function of metallicity in log space, assuming Gaussian probability distributions. For non-polluted sources, we compare results from three different probability distributions. First, we take the non-polluted white dwarfs to have mean accretion rates of 0 g s$^{-1}$ and 3$\sigma$ uncertainties corresponding to their upper limits $R_{\mathrm{NP}}$, resulting in the following likelihood $\mathcal{L}$:
\begin{equation}
\mathcal{L} = \prod_{i=0}^{N_{\mathrm{P}}} \mathcal{N}(R_{\mathrm{P},i},\sigma_{\mathrm{P},i}) \times \prod_{j=0}^{N_{\mathrm{NP}}}\mathcal{N}(0,R_{\mathrm{NP},j})
\end{equation}
where $N_{\mathrm{P}}$ and $N_{\mathrm{NP}}$ are defined as the polluted and non-polluted sample sizes, respectively. We additionally require the predicted accretion rates to be non-negative. We calculate a best-fit slope of ($-$7$\pm$9) $\times$ 10$^5$ g s$^{-1}$ dex$^{-1}$, consistent (within 1$\sigma$) with a slope of 0 g s$^{-1}$ dex$^{-1}$. Our inferred accretion rates, corresponding Gaussian probability distributions, and best-fit slope are shown in Figure~\ref{accretion2}. We note that we would expect a positive, rather than negative, slope if giant planets are the main drivers of accretion.}

\bedit{To test the robustness of this result, we also fit the accretion rate upper limits using two other probability distributions: log uniform and sigmoid distributions.  As with the Gaussian probability distributions, we require the predicted accretion rates to be non-negative. The log uniform distribution additionally requires them to be less than the upper limit. The sigmoid distribution includes a fourth parameter $k$ that determines the steepness of the sigmoid slope, resulting in the following likelihood function: 
\begin{equation}
\mathcal{L} = \prod_{i=0}^{N_{\mathrm{P}}} \mathcal{N}(R_{\mathrm{P},i},\sigma_{\mathrm{P},i}) \times \prod_{j=0}^{N_{\mathrm{NP}}}\frac{1}{f(1+e^{-k(x_j - R_{\mathrm{NP},j})})}
\end{equation}
where $f$ is a normalization function. We fit the non-polluted targets in log space and define $f$ as the integral of the sigmoid from $x=-$20 to 20. We require $k$ to be $<-$1 to prohibit arbitrarily shallow sigmoid distributions. We find best-fit slopes of ($-$1.2$\pm$1.4) $\times$ 10$^6$ and ($-$1.2$\pm$1.8) $\times$ 10$^6$ g s$^{-1}$ dex$^{-1}$ for the log uniform and sigmoid probability distributions, respectively. Both are consistent with a slope of 0 g s$^{-1}$ dex$^{-1}$. We therefore find no significant association between the metallicities and inferred accretion rates. }

\subsection{Alternate Metallicity Dependencies as Potential Sources of Bias}
\label{sec:relations}
Stellar metallicity is known to be correlated with many parameters\bedit{, such as age \citep[e.g.,][]{twa1980,sou2008}, binary separation \citep[e.g.,][]{elb2019}, and location within the galaxy \citep[e.g.,][]{Ber2014,lia2023}}. The metallicity distributions in Figure~\ref{met_distr} may therefore include contributions from additional metallicity dependencies. Relevant to our study are the age-metallicity and separation-metallicity relations. To determine whether these may be contributing to our observed metallicity distributions, we investigate their impact in further detail below. 

\subsubsection{Age-Metallicity Relation}
\label{sec:agerelation}
The polluted and non-polluted systems have different age distributions, as seen in the right panel of Figure~\ref{characteristics}. Because age is often correlated with metallicity, we investigate whether this could be introducing biases to the metallicity measurements shown in Figure~\ref{met_distr}. 

We first calculate each white dwarf's cooling age using the most reliable literature values for the effective temperature, surface gravity, and spectral class. \bedit{When available, we use parameters derived from photometric fits.} We then use the cooling tracks from \citet{ber2020} to calculate the age. We do not calculate cooling ages for two stars due to insufficient {\it Gaia} data (TIC 176879526) and a lack of available atmospheric parameters (TIC 702485620). We are unable to obtain reliable uncertainties for three additional targets (TICs 630254489, 435915513, and 76956031), and therefore do not report their cooling ages in Table~\ref{tab:targetlist}. 

 However, the white dwarfs' cooling ages do not necessarily represent the distribution of the binary systems' true ages. To estimate the total age of a given system, we use methods similar to those outlined by \citet{man2016}. We first estimate each white dwarf's progenitor mass using the initial-to-final mass relation from \citet{cat2008}. We note that this relation was derived exclusively for \bedit{DA white dwarfs.} However, we do not expect this to have a significant effect on the results, as demonstrated by \citet{man2016}. We then use MESA Isochrones and Stellar Tracks \citep[MIST,][]{dot2016, cho2016} to estimate the progenitor lifetime given the stellar metallicity and calculated mass. This lifetime is added to the cooling age of the white dwarf to determine the total age of the binary system. We only perform this calculation for white dwarfs with masses above \bedit{$0.63$ $M_{\odot}$.} This selection is intended to exclude systems that have undergone common-envelope evolution or mass transfer, and where the initial-to-final mass relation is therefore not applicable. \bedit{Additionally, \citet{hei2022} found that age estimates are most reliable in this regime.} Our final sample of white dwarfs with estimated ages includes \bedit{14} targets. 

\citet{man2021} performed a similar calculation to estimate the local age-metallicity relation. Using 46 white dwarf binary systems found with SDSS and 189 proper motion pairs found with \textit{Gaia} DR2, with ages spanning over 10 Gyr, the authors did not find strong evidence for a correlation between age and metallicity. We similarly do not find a significant correlation, as shown in the right panel of Figure~\ref{characteristics}, where the black markers denote the average metallicity for each 1 Gyr bin. For bins with at least three systems, we calculate the standard deviation from the observed spread in metallicities. For bins with fewer systems, we combine the metallicity uncertainties of individual measurements ($0.08$ dex). We observe no significant trend in metallicity as a function of age. Given that the initial-to-final mass relation for white dwarfs is poorly constrained, we  also derive the total ages using the initial-to-final mass relations from \citet{ges2014} and \citet{cum2018}. Using these relations, we still find no evidence for a significant trend between age and metallicity. Additionally, when we fit a linear model to our individual metallicity measurements \bedit{using an MCMC sampler}, we find a slope of \bedit{$-$0.02 $\pm$ 0.04,} consistent with there being no correlation. We conclude that the systems' ages do not bias our results. 

\bedit{We note that 12 of the 14 white dwarfs with estimated ages have atmospheric parameters obtained from photometric fits. While there are known discrepancies between parameters obtained with photometry and spectroscopy \citep[e.g.,][]{gen2019}, our conclusions are not changed when we exclude the remaining two stars (TICs 429102184 and 352817378) from our sample. }

\subsubsection{Separation-Metallicity Relation}
\label{sec:separationrelation}
The observed metallicity distributions may also include effects from the semi-major axis separations. \citet{elb2019} searched for a relation between orbital separation and metallicity for binary systems in \textit{Gaia} DR2. They found that a dependence between separation and metallicity emerges for separations below $\sim$250 \bedit{au}. This is possibly driven by the different formation mechanisms for close and wide binaries---fragmentation of unstable disks can create close binaries, while core fragmentation can create wide binaries \citep[e.g.,][]{moe2019}. Our sample contains \bedit{four} systems with separations below this threshold, including one polluted white dwarf system. 

To investigate whether orbital separation is contributing to the observed metallicity distributions, we repeat our analysis using only systems with separations greater than 250 \bedit{au}. Our final mean metallicities for polluted and non-polluted systems are consistent with the values previously derived. We calculate a mean and standard deviation of $-0.21\pm0.05$ and $0.21\pm0.04$ dex for the polluted systems, and $-0.17\pm0.04$ and $0.24\pm0.03$ dex for the non-polluted systems. These are consistent with our previous results within uncertainties. The separation-metallicity relation therefore does not have a significant effect on our results.

\begin{table*}
\fontsize{7}{10}\selectfont
\centering
\caption{List of TIC IDs for 59 white dwarf binaries. \bedit{We also indicate the binary separations, in addition to the white dwarfs' effective temperatures, \bedittwo{surface gravities,} cooling ages, total ages (see \S\ref{sec:agerelation}), and spectral classifications.} Polluted white dwarfs are denoted by a `Z' in their spectral type. \bedit{We cite the corresponding sources for the atmospheric parameters and spectral classifications.} } \label{tab:targetlist}
\centering
\begin{tabular}{ccccccccc}
\hline
\hline
 Companion ID & White Dwarf ID & \bedit{Projected} Separation & T$_{\textrm{eff}}$ & \bedittwo{log($g$)} & Cooling Age & Total Age &  WD Type & References \\
 &  &  (\bedit{au}) & (K) & & (Gyr)  & (Gyr) & & \\
\hline
 376455973 & 611128637 & 1431.4 & 5780$\pm$230 & 7.84$\pm$0.20 & 2.49$^{+1.22}_{-0.64}$ &  -- & DZ & 1, 2 \\ 
 191145234 & 191145238 & 2595.5 & 10010$\pm$50 & 7.97$\pm$0.01 & 0.61$^{+0.01}_{-0.01}$ &  -- & DA & 3, 4 \\ 
 268127217 & 611261822 & 4990.9 & 4760$\pm$30 & 7.74$\pm$0.02 & 3.39$^{+0.35}_{-0.13}$ &  -- & DZAH & 3, 3 \\ 
 336892483 & 610972154 & 1940.7 & 19350$\pm$1060 & 8.09$\pm$0.08 & 0.09$^{+0.03}_{-0.02}$ & 0.78$^{+0.62}_{-0.29}$ & DAZ & 5, 5 \\ 
 35703955 & 630323435 & 1377.5 & 4820$\pm$130 & 7.62$\pm$0.22 & 3.15$^{+1.44}_{-0.75}$ &  -- & DZ & 1, 6 \\ 
 270371546 & 630406417 & 25550.8 & 8580$\pm$470 & 7.62$\pm$0.57 & 0.70$^{+0.68}_{-0.27}$ &  -- & DZ & 1, 2 \\ 
 129904995 & 129905006 & 1548.1 & 5980$\pm$30 & 8.00$\pm$0.01 & 2.37$^{+0.04}_{-0.04}$ &  -- & DA & 3, 7 \\ 
 251083146 & 630254489 & 928.8 & -- & -- & -- & --& DB & 8 \\ 
 422890692 & 630273894 & 1169.9 & 5810$\pm$1210 & 7.87$\pm$0.95 & 3.54$^{+4.15}_{-2.33}$ &  -- & DZ & 5, 9 \\ 
 439869953 & 439869954 & 1298.5 & 7890$\pm$60 & 7.97$\pm$0.01 & 1.11$^{+0.03}_{-0.02}$ &  -- & DA & 10, 11 \\ 
 381347690 & 640215257 & 1306.6 & 4100$\pm$50 & 7.73$\pm$0.04 & 5.09$^{+0.54}_{-0.33}$ &  -- & DC & 3, 3 \\ 
 279049424 & 649756572 & 3160.8 & 6460$\pm$130 & 7.83$\pm$0.07 & 1.61$^{+0.21}_{-0.16}$ &  -- & DA & 5, 12 \\ 
 435915513 & 435915513 & 60.5 & -- & -- & -- & --& DA & 13 \\ 
 429102184 & 429102184 & 15.8 & 16420$\pm$420 & 8.46$\pm$0.20 & 0.30$^{+0.14}_{-0.12}$ & 0.47$^{+0.20}_{-0.14}$ & DA & 14, 14 \\ 
 445709402 & 445709423 & 2326.1 & 8690$\pm$140 & 7.92$\pm$0.03 & 0.83$^{+0.04}_{-0.04}$ &  -- & DA & 10, 10 \\ 
 125838647 & 427771898 & 2226.5 & 5560$\pm$30 & 8.05$\pm$0.02 & 3.49$^{+0.24}_{-0.24}$ &  -- & DA & 15, 11 \\ 
 91690128 & 91690158 & 4824.1 & 11340$\pm$230 & 8.05$\pm$0.04 & 0.50$^{+0.04}_{-0.04}$ &  -- & DB & 5, 5 \\ 
 53088536 & 702485620 & 1616.0 & -- & -- & -- & --& DA & 12 \\ 
 171848237 & 703420509 & 2060.3 & 6690$\pm$40 & 8.04$\pm$0.01 & 1.88$^{+0.04}_{-0.04}$ &  -- & DA & 10, 10 \\ 
 172703279 & 172703276 & 508.5 & 9080$\pm$60 & 7.90$\pm$0.01 & 0.72$^{+0.01}_{-0.01}$ &  -- & DA & 3, 11 \\ 
 417756881 & 743509720 & 2445.0 & 4980$\pm$30 & 7.53$\pm$0.03 & 2.49$^{+0.13}_{-0.09}$ &  -- & DC & 3, 12 \\ 
 280310048 & 471011543 & 15.1 & 7590$\pm$40 & 7.96$\pm$0.02 & 1.27$^{+0.04}_{-0.03}$ &  -- & DQZ & 1, 16 \\ 
 293414051 & 800494479 & 1221.4 & 6570$\pm$320 & 8.09$\pm$0.15 & 2.15$^{+0.71}_{-0.38}$ & 3.17$^{+3.87}_{-0.76}$ & DAZ & 5, 5 \\ 
 241194061 & 241194059 & 1415.4 & 11670$\pm$90 & 7.74$\pm$0.01 & 0.30$^{+0.01}_{-0.01}$ &  -- & DA & 10, 10 \\ 
 1732730 & 841989747 & 4729.4 & 9480$\pm$60 & 8.15$\pm$0.02 & 0.89$^{+0.02}_{-0.02}$ & 1.58$^{+0.18}_{-0.14}$ & DAZ & 3, 6 \\ 
 374210257 & 842251290 & 6472.1 & 7760$\pm$570 & 7.96$\pm$0.26 & 1.37$^{+0.48}_{-0.28}$ &  -- & DA & 5, 5 \\ 
 17301096 & 900244294 & 14110.4 & 7210$\pm$150 & 8.34$\pm$0.30 & 1.89$^{+1.24}_{-0.32}$ & 2.14$^{+1.48}_{-0.35}$ & DZ & 1, 6 \\ 
 47350711 & 47353019 & 2778.3 & 11160$\pm$410 & 7.67$\pm$0.09 & 0.31$^{+0.05}_{-0.05}$ &  -- & DA & 5, 2 \\ 
 20324932 & 20324921 & 4016.8 & 5960$\pm$40 & 8.17$\pm$0.01 & 3.98$^{+0.14}_{-0.14}$ & 4.58$^{+0.23}_{-0.19}$ & DC & 3, 12 \\ 
 239209564 & 903441740 & 3982.9 & 8930$\pm$360 & 8.04$\pm$0.13 & 0.96$^{+0.16}_{-0.12}$ &  -- & DAZ & 1, 1 \\ 
 138545439 & 900545886 & 2501.6 & 9510$\pm$480 & 8.16$\pm$0.13 & 0.88$^{+0.19}_{-0.14}$ & 1.54$^{+1.11}_{-0.40}$ & DZ & 5, 17 \\ 
 176879529 & 176879526 & 9186.2 & -- & -- & -- & --& DA & 11 \\ 
 347389665 & 347389664 & 974.4 & 10430$\pm$80 & 8.02$\pm$0.01 & 0.58$^{+0.01}_{-0.01}$ &  -- & DA & 10, 10 \\ 
 458455132 & 166668903 & 5449.1 & 14510$\pm$250 & 7.96$\pm$0.01 & 0.21$^{+0.01}_{-0.01}$ &  -- & DA & 10, 7 \\ 
 445863106 & 1001309484 & 8670.3 & 6980$\pm$260 & 8.14$\pm$0.19 & 1.94$^{+0.64}_{-0.29}$ & 2.67$^{+4.22}_{-0.58}$ & DZ & 1, 2 \\ 
 284628177 & 284628173 & 3476.4 & 14490$\pm$350 & 7.95$\pm$0.08 & 0.23$^{+0.03}_{-0.02}$ &  -- & DBAZ & 18, 11 \\ 
 420807841 & 1100360772 & 5472.2 & 6330$\pm$190 & 8.03$\pm$0.10 & 2.48$^{+0.56}_{-0.37}$ &  -- & DZ & 5, 3 \\ 
 229874145 & 1101440031 & 1389.8 & 5210$\pm$40 & 7.83$\pm$0.02 & 3.11$^{+0.28}_{-0.20}$ &  -- & DA & 10, 10 \\ 
 160424250 & 1102569055 & 1120.7 & 5650$\pm$80 & 7.49$\pm$0.03 & 1.76$^{+0.08}_{-0.12}$ &  -- & DC & 3, 19 \\ 
 149692037 & 149692034 & 3219.3 & 16940$\pm$140 & 7.89$\pm$0.01 & 0.11$^{+0.00}_{-0.00}$ &  -- & DA & 10, 11 \\ 
 207468713 & 1201118897 & 814.2 & 4900$\pm$370 & 8.04$\pm$0.27 & 6.67$^{+1.86}_{-1.96}$ &  -- & DAZ & 5, 20 \\ 
 318811655 & 1204769581 & 12307.9 & 15260$\pm$340 & 7.91$\pm$0.04 & 0.17$^{+0.02}_{-0.02}$ &  -- & DA & 5, 5 \\ 
 5876391 & 5876393 & 527.2 & 4860$\pm$60 & 7.16$\pm$0.04 & 1.81$^{+0.06}_{-0.14}$ &  -- & DC & 3, 21 \\ 
 162690651 & 1201039482 & 3145.1 & 5260$\pm$30 & 8.10$\pm$0.02 & 5.54$^{+0.30}_{-0.30}$ & 6.51$^{+0.43}_{-0.38}$ & DAZ & 22, 23 \\ 
 21860383 & 21860382 & 2451.7 & 12030$\pm$100 & 8.00$\pm$0.01 & 0.39$^{+0.01}_{-0.01}$ &  -- & DA & 10, 10\\ 
 219857965 & 1271265755 & 1530.6 & 6530$\pm$20 & 7.93$\pm$0.01 & 1.69$^{+0.02}_{-0.02}$ &  -- & DA & 3, 7 \\ 
 377626776 & 1505419059 & 799.2 & 22330$\pm$2450 & 8.06$\pm$0.14 & 0.06$^{+0.05}_{-0.03}$ & 0.68$^{+1.07}_{-0.34}$ & DA & 5, 24 \\ 
 277255071 & 277255078 & 1214.2 & 9710$\pm$90 & 7.99$\pm$0.01 & 0.68$^{+0.02}_{-0.02}$ &  -- & DA & 3, 4 \\ 
 23395123 & 23395129 & 4708.0 & 6520$\pm$90 & 7.84$\pm$0.05 & 1.60$^{+0.15}_{-0.12}$ &  -- & DA & 5, 5\\ 
 229770891 & 229770890 & 2727.8 & 7210$\pm$160 & 7.51$\pm$0.06 & 0.88$^{+0.08}_{-0.07}$ &  -- & DA & 5, 3 \\ 
 76956031 & 76956031 & 1.6 & -- & -- & -- & --& DA & 8 \\ 
 352817358 & 352817378 & 5701.4 & 14390$\pm$240 & 8.27$\pm$0.10 & 0.33$^{+0.08}_{-0.05}$ & 0.63$^{+0.20}_{-0.11}$ & DB & 25, 11 \\ 
 305267214 & 2000305088 & 3931.2 & 11650$\pm$530 & 8.13$\pm$0.11 & 0.51$^{+0.10}_{-0.08}$ & 1.34$^{+1.08}_{-0.45}$ & DAZ & 1, 2 \\ 
 259014661 & 353035668 & 4090.6 & 4680$\pm$80 & 7.13$\pm$0.09 & 1.94$^{+0.14}_{-0.22}$ &  -- & DC & 5, 26 \\ 
 36726370 & 36726368 & 1536.8 & 6690$\pm$60 & 7.98$\pm$0.02 & 1.72$^{+0.06}_{-0.06}$ &  -- & DA & 15, 11 \\ 
 427863040 & 427863042 & 894.7 & 9560$\pm$40 & 8.07$\pm$0.01 & 0.78$^{+0.01}_{-0.01}$ & 1.66$^{+0.25}_{-0.19}$ & DA & 3, 27 \\ 
 428986692 & 428986688 & 861.6 & 12380$\pm$250 & 8.30$\pm$0.03 & 0.58$^{+0.04}_{-0.04}$ & 0.92$^{+0.10}_{-0.07}$ & DC & 5, 28 \\ 
 91102009 & 2053962932 & 2037.8 & 11930$\pm$130 & 8.01$\pm$0.02 & 0.41$^{+0.02}_{-0.02}$ &  -- & DA & 5, 5 \\ 
 9726029 & 9726026 & 4500.0 & 18840$\pm$360 & 7.98$\pm$0.06 & 0.08$^{+0.01}_{-0.01}$ &  -- & DA & 25, 7 \\ 

\hline
\multicolumn{8}{l}{%
  \begin{minipage}{15cm}%
   \bedit{References: (1) \citet{cou2019}; (2) \citet{kle2013};  (3) \citet{car2023}; (4) \citet{bas2003}; (5) \citet{fus2019}; (6) \citet{kep2015}; (7) \citet{gia2011}; (8) \citet{hol2013}; (9) \citet{mar2023}; (10) \citet{kil2020}; (11) \citet{zuc2014}; (12) \citet{dup1994}; (13) \citet{joy2018}; (14) \citet{lan1996}; (15) \citet{blo2019}; (16) \citet{hol2008}; (17) \citet{tre2017}; (18) \citet{ber2011}; (19) \citet{osw1994}; (20) \citet{kon2021}; (21) \citet{rei2005}; (22) \citet{blo2022}; (23) \citet{kil2006}; (24) \citet{hol2003}; (25) \citet{bed2017}; (26) \citet{lim2015}; (27) \citet{rei1996}; (28) \citet{osw1988}} %
  \end{minipage}%
}\\
\end{tabular}
\end{table*}

\section{Experimental Caveats}
\label{sec:caveats}
\begin{figure}
\centering
\includegraphics[width=\columnwidth]{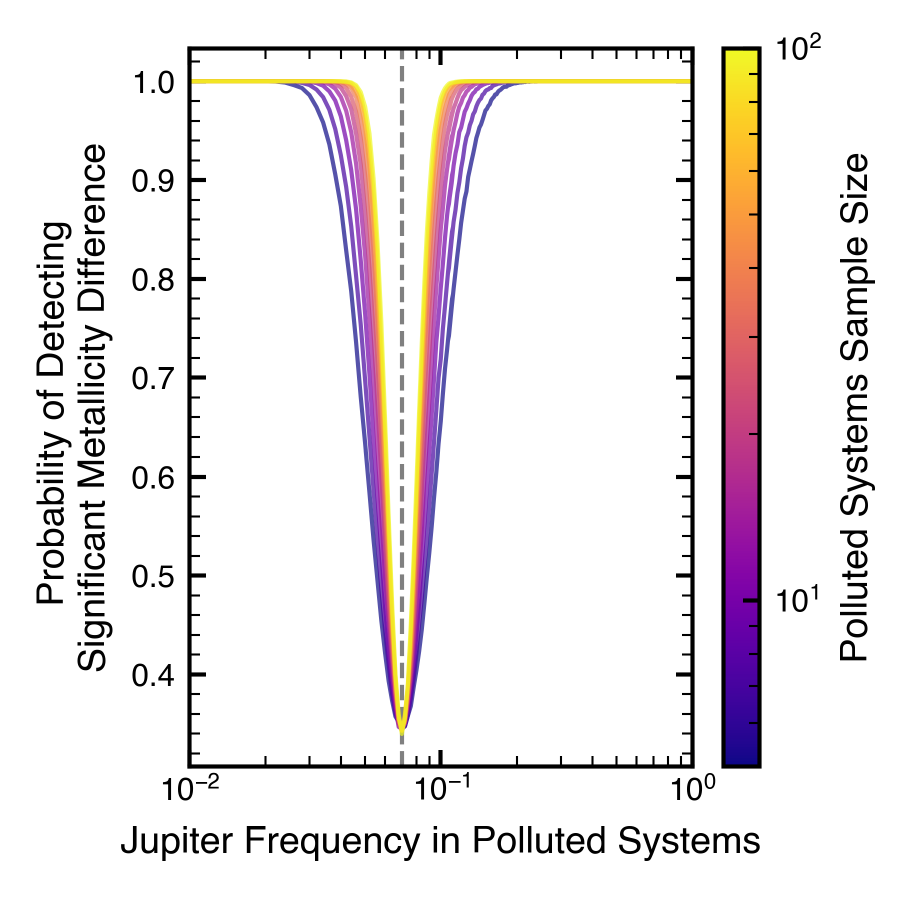}
\caption{Sensitivity of sample as a function of Jupiter frequencies in polluted systems. Sensitivity is measured as the fraction of samples where we would expect to observe a significant difference in the mean metallicities of polluted and non-polluted systems. We assume a constant Jupiter frequency in non-polluted systems of $0.07$, consistent with our observed value. This value is denoted by the vertical dashed line, while sensitivities for various samples sizes are denoted by the colored curves. Future work could identify a larger sample of polluted white dwarfs in binary systems. This would increase the probability of detecting a significant metallicity offset if polluted and non-polluted systems have different Jupiter frequencies. \label{jup_freq_samples}}
\end{figure}

Although our observations suggest that Jupiter-mass planets are not necessarily the dominant driver of white dwarf pollution, several factors complicate this interpretation. We note \bedit{four} limitations to our analysis, and suggest ways to mitigate them in the future. 
\begin{description}

\item[$\bullet$ \textbf{Non-uniform pollution sensitivity}] 
\bedit{As discussed in \S\ref{sec:accretion}, it is significantly easier to detect pollution signatures in He-dominated atmospheres than in H-dominated atmospheres. The discrepancy between the number of He-dominated white dwarfs in the polluted \bedittwo{(12 out of 18)} and non-polluted \bedittwo{(7 out of 41)} samples likely introduces a bias in our spectral classifications. In other words, some of the H-dominated white dwarfs in our non-polluted sample may be misclassified because they have not been observed with sufficient sensitivity. If polluted white dwarf systems do have higher metallicities, then misclassified objects would shift the metallicity distribution of non-polluted systems to higher metallicities, making it more difficult to differentiate between polluted and non-polluted distributions. To account for this, we also compare inferred accretion rates (for polluted targets) and upper bounds (for non-polluted targets) in \S\ref{sec:accretion}. }

  \item[$\bullet$ \textbf{Non-uniform spectra quality}] The available spectra for our sample of white dwarfs is non-uniform, as the white dwarfs have been observed with different instruments at different \bedit{SNRs}. Because the quality of the available spectra vary, it is difficult to perform a consistent classification of white dwarf spectral classes. Therefore, some of the white dwarfs in our sample may be incorrectly categorized as non-polluted because they were not observed with sufficient SNR in the appropriate wavelength regimes. \bedit{For instance, only 3 of the 41 non-polluted systems have been observed with the {\it Hubble Space Telescope} ({\it HST}) in the ultraviolet, a wavelength regime especially sensitive to pollution.} Because detecting signatures of pollution implies that the target was observed with sufficient SNR to properly classify it, we consider it unlikely that the polluted white dwarfs are misclassified. 

    \item[] To investigate any biases introduced by non-uniform spectra quality, we restrict our sample of non-polluted systems to \bedit{13} targets that have high-resolution spectroscopy available. We are therefore more confident in their spectral classification. \bedit{Six} of these systems are included in \citet{zuc2014}. Using this updated sample, we estimate the mean metallicity to be \bedit{$-0.17\pm0.09$ dex}, with a standard deviation of \bedit{$0.31\pm0.08$ dex.} This is consistent with the original metallicity estimations for both our polluted and non-polluted samples. We therefore conclude that our results are not significantly biased by the lack of high-resolution spectra for \bedit{the remaining 28} non-polluted targets.

  \item[$\bullet$ \textbf{Small sample size}] Due to the limited number of known polluted white dwarfs in binary systems, we are only able to observe 18 such systems. This small sample size restricts how sensitive our results are to differences in mean metallicities. The effect of our sample size is demonstrated in Figure~\ref{jup_freq_samples}--as the number of polluted white dwarf systems increases, the probability of detecting a significant metallicity difference for a given Jupiter frequency increases. Future work using an expanded sample of polluted systems would therefore be more sensitive to any offsets between the polluted and non-polluted metallicity distributions. For our sample size, we would expect nearly all polluted samples to exhibit a significant difference in metallicities from the non-polluted sample if the polluted and non-polluted Jupiter frequencies are $1.0$ and $0.11$, respectively. 

    \item[$\bullet$ \textbf{Selection of binary systems}] Our interpretation is also limited by our target selection process, which requires that all white dwarfs are in binary systems. This likely biases our sample, as specific pollution mechanisms may dominate among binary systems. For instance, effects such as the Eccentric Kozai-Lidov mechanism \citep{ste2017} may be contributing to the observed pollution. This would suppress any contributions from massive planets. To avoid these potential biases, future work could use alternative approaches to estimate the polluted white dwarfs' primordial metallicities. Such an estimation is possible using nearby stars that formed from the same interstellar gas as the white dwarf, though this would not necessarily have as strong dynamical consequences as a visual binary
    . This criterion is satisfied by stars in open clusters \citep{bov2016, cas2020}. Additionally, stars in the same moving group tend to be closer in metallicity than average field stars, though they can still have a large metallicity dispersion \citep[$\sim$0.2 \bedit{-} 0.3 dex, e.g.,][]{ant2008}. Extending the analysis performed in \S\ref{sec:Results} to a sample of single-star polluted and non-polluted white dwarf pairs with a shared formation history would provide further constraints on the mechanisms driving white dwarf pollution. 
\end{description}

\section{\bedit{Prospects for Future Study}}
\label{sec:future}
The caveats described in \S\ref{sec:caveats} limit our sensitivity and the interpretation of our results. However, given the current available data, we do not detect a significant difference in the metallicity distributions of polluted and non-polluted systems. \bedit{Here, we describe prospects for future study, with a focus on addressing the first three experimental caveats and discussing potential estimates of Jupiter frequency.}

\subsection{Sample Characteristics}
\label{newsample}
\bedit{Future work could provide uniform spectral classifications and an expanded sample size by selecting and observing a curated sample of white dwarfs in binaries. This sample would ideally have the following characteristics:}
\begin{description}
\item[$\bullet$ \textbf{Uniform pollution sensitivity}] \bedit{As discussed in \S\ref{sec:accretion}, it is easier to detect pollution in He-dominated atmospheres than in H-dominated atmospheres. To ensure that observations of polluted and non-polluted targets are similarly sensitive to pollution signatures, a future sample could be limited to He-dominated white dwarfs.}
\item[$\bullet$ \textbf{Uniform spectra quality}] \bedit{To avoid discrepancies in the quality of spectra, the sample of white dwarfs could be uniformly observed with the same instrument setup. This would ensure that all spectra have the same spectral ranges and resolutions, in addition to the same flux calibrations and similar SNRs. These observations would enable a more fair comparison of polluted and non-polluted targets. }

\item[$\bullet$ \textbf{Larger sample size}] \bedit{Our sample includes 59 white dwarfs. However, cross-matching known white dwarfs in binaries \citep{fus2019, elb2021} with the Large Sky Area Multi-Object Fibre Spectroscopic Telescope \citep[LAMOST, ][]{den2012, zha2012, liu2013} DR8, RAdial Velocity Experiment \citep[RAVE, ][]{cas2017}, and GALactic Archaeology with HERMES \citep[GALAH, ][]{bud2021} DR3 catalogs provides $\sim$1,300 binary systems with measured companion metallicities. $\sim$400 of these systems have white dwarfs with {\it G}-band magnitude $<$ 19. It would therefore be possible to obtain a significantly larger sample of polluted and non-polluted white dwarfs in binary systems. This would increase sensitivity to any differences in the metallicity distributions of both populations. }
\end{description}

\subsection{Expected Metallicity Difference for Jupiter Systems}
\label{sec:prediction_comparison}
\bedit{The results of planet search surveys can be used to predict the} expected difference in metallicity between polluted and non-polluted white dwarf systems if Jupiter-like planets are a dominant cause of white dwarf pollution. 
\bedit{The California Legacy Survey \citep[CLS,][]{lee2021} catalog, for example, includes 719 FGKM stars with 178 detected planets, and was compiled without any bias toward stars with planets. Though it may not be} possible to firmly put \bedit{current or future} metallicity measurements on the same absolute footing as the CLS sample \bedit{(see \S\ref{sec:field_comparison}),} it is still instructive to use their results to understand the offset in metallicity between stars without Jupiter-like planets and stars hosting cold Jupiters. \citet{buc2018} showed that the planet/metallicity correlation is strongest for eccentric or close-in Jupiters, possibly because higher metallicity systems tend to host multiple massive planets, leading to more planet-planet interactions. However, close-in planets are not likely to survive red giant evolution. We are therefore primarily interested in the metallicity offset between systems hosting a cold Jupiter and systems with no known Jupiter analogues.

\begin{figure}
\centering
\includegraphics[width=\columnwidth]{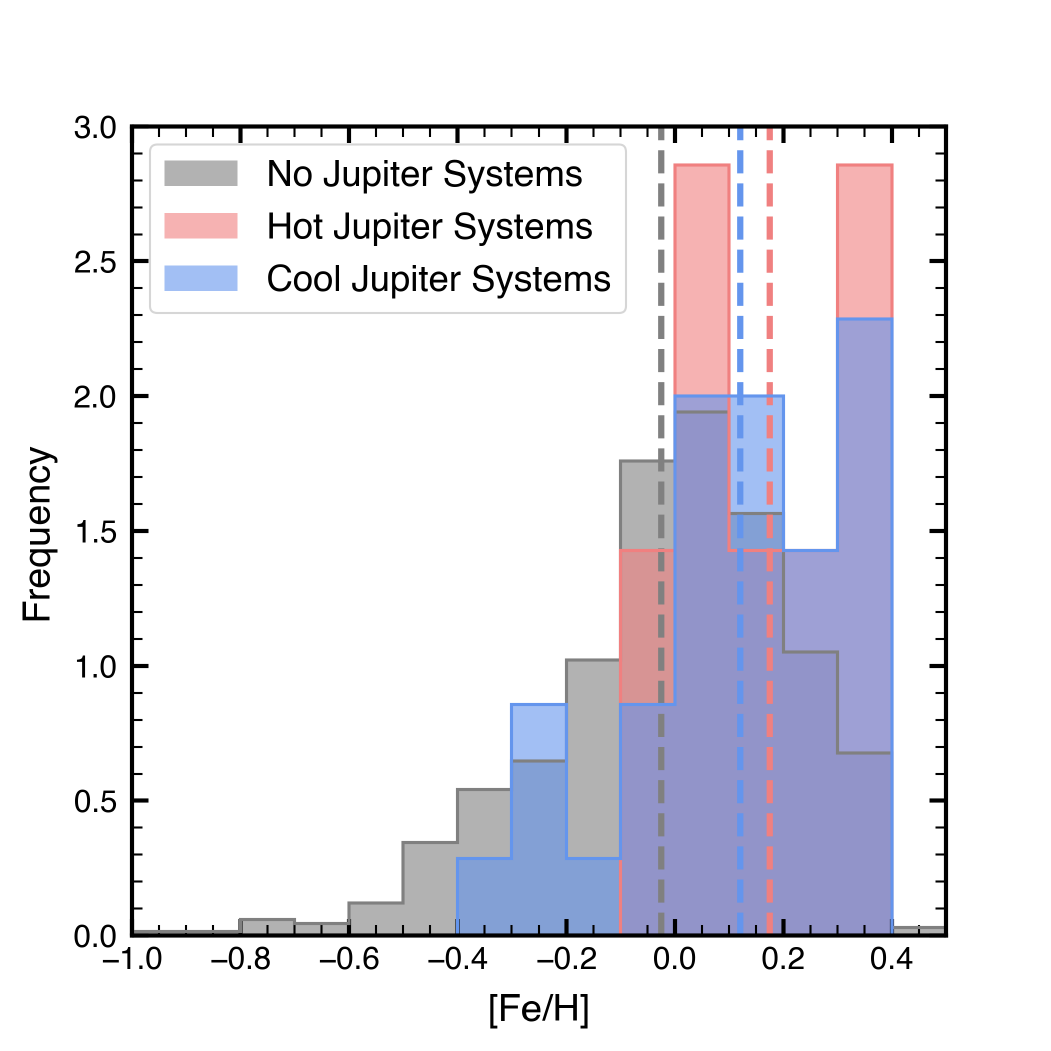}
\caption{Normalized histogram showing the metallicities of systems with hot (pink), cool (blue) and no (gray) Jupiter analogues in the California Legacy Survey. Dashed lines show the mean metallicity for each sample. Selection criteria are described in \S\ref{sec:prediction_comparison}. Though systems with cool Jupiters tend to have lower metallicities than those with hot Jupiters, consistent with the results from \citet{buc2018}, they still have a higher mean metallicity than systems with no Jupiter anologues. We therefore expect to observe a metallicity offset between polluted and non-polluted white dwarf systems if cool Jupiters are more common around polluted white dwarfs.    \label{CoolJup}}
\end{figure}

Because \citet{buc2018} do not consider systems without Jupiters, we use CLS \citep{lee2021} to probe these metallicity distributions.  
We apply the selection criteria implemented by \citet{buc2018}: planets with a mass between 0.3 to 3.0 $M_{\textrm{J}}$ are considered Jupiter analogues, Jupiters with eccentricities greater than 0.25 are classified as eccentric, and Jupiters with a semi-major axis less than 0.1 \bedit{au} are classified as hot Jupiters. In total, we identify 35 cool Jupiters, an additional 12 cool eccentric Jupiters, and 7 hot Jupiters. 
We calculate the mean metallicity and standard deviations for systems without Jupiters ($\mu_{\textrm{NJ}}=-0.03\pm0.01$, $\sigma_{\textrm{NJ}}=0.29\pm0.01$), systems with cool non-eccentric Jupiters ($\mu_{\textrm{J}}=0.12\pm0.04$, $\sigma_{\textrm{J}}=0.19\pm0.07$), systems with cool eccentric Jupiters ($\mu_{\textrm{EJ}}=0.18\pm0.04$, $\sigma_{\textrm{EJ}}=0.15\pm0.04$), and systems with hot Jupiters ($\mu_{\textrm{HJ}}=0.17\pm0.07$, $\sigma_{\textrm{HJ}}=0.19\pm0.07$). The metallicity distributions for no Jupiter, hot Jupiter, and cool Jupiter systems are shown in Figure~\ref{CoolJup}. We note that, while we observe similar overall trends, our mean metallicities for hot, eccentric, and cool Jupiter systems are lower than the \citet{buc2018} values ($0.25\pm0.03$, $0.23\pm0.04$, and $-0.07\pm0.05$, respectively). These differences may be the result of sample selection biases. 

These results demonstrate that, while cool Jupiter systems tend to have a lower metallicity than hot and eccentric Jupiter systems, they still tend to have a higher metallicity than systems with no known Jupiters. \bedit{We therefore expect to observe a $\sim$0.15  $\pm$ 0.04 dex metallicity offset between our sample of polluted and non-polluted white dwarf systems if such planets are the dominant contributors to pollution.}

\subsection{Predicted Jupiter Frequency}
\begin{figure}
\centering
\includegraphics[width=\columnwidth]{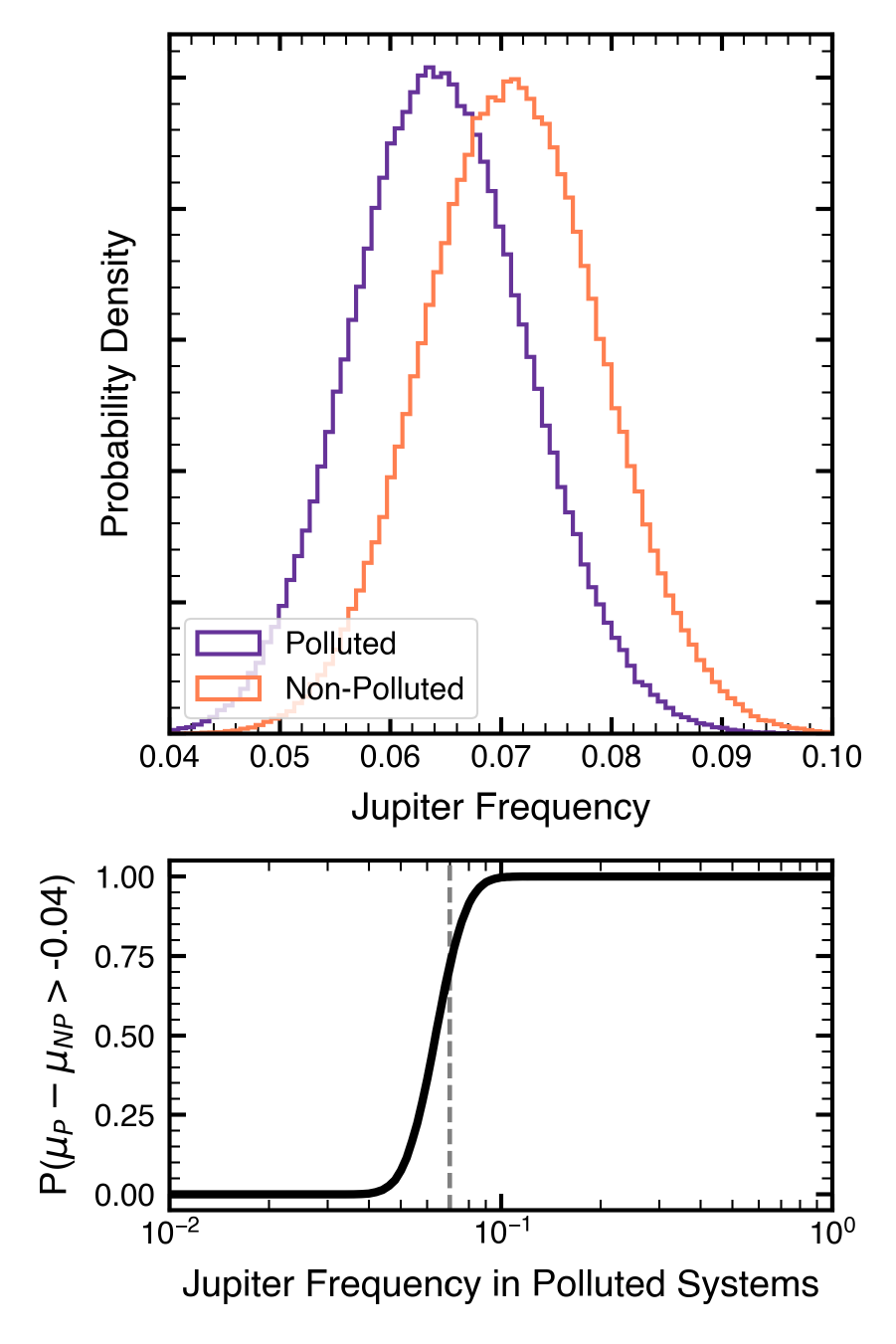}
\caption{Top: Predicted Jupiter frequencies for polluted and non-polluted systems, shown in purple and orange, respectively. Jupiter frequencies are calculated using Equation~\ref{eq:power_law}. We do not identify a significant difference between the two distributions. Bottom: Probability of finding a metallicity difference ($\mu_{\textrm{P}} - \mu_{\textrm{NP}}$) greater than the one we observe ($-0.04$) as a function of the Jupiter frequency in polluted systems. We assume a constant Jupiter frequency in non-polluted systems of $0.07$, consistent with our observed value. This value is denoted by the vertical dashed line. \label{jup_freq}}
\end{figure}

\bedit{Given a sample as described in \S\ref{newsample}, it would be possible to} estimate the relative fraction of Jupiter analogs in polluted and non-polluted systems. We \bedit{could} use the giant planet occurrence rate provided by \citet{ghe2018} to estimate Jupiter frequencies. The authors use a spectroscopic survey of 245 subgiants, in addition to a sample of previously observed main-sequence FGKM stars from \citet{joh2010}, to estimate the giant planet occurrence rate $f$ as a function of the stellar mass $M_{\star}$ and metallicity $F$:
\begin{equation}
\label{eq:power_law}
f(M_{\star},F) = 0.08^{+0.008}_{-0.010}(M_{\star}\times10^F)^{1.04^{+0.12}_{-0.16}}
\end{equation}
This result is tested for stellar masses up to $2$ M$_{\odot}$ and stellar metallicities ranging from approximately $-2.0$ to $0.5$. We use Equation~\ref{eq:power_law} to estimate the ratio of Jupiter frequencies in polluted and non-polluted systems. Because we are only interested in the ratio of frequencies as a function of metallicity, we neglect the stellar mass term. We then calculate the mean frequency for polluted systems $f_{\textrm{P}}$ and for non-polluted systems $f_{\textrm{NP}}$. \bedit{For our sample, we} calculate a ratio of $f_{\textrm{P}}/f_{\textrm{NP}}$ $=$ 0.91$^{+0.10}_{-0.08}$, consistent with a value of one. These results for polluted and non-polluted systems are shown in the top panel of Figure~\ref{jup_freq}. We repeat this calculation restricting our sample to the 10 polluted and 33 non-polluted systems with white dwarf progenitor masses less than $2$ M$_{\odot}$. We calculate $f_{\textrm{P}}/f_{\textrm{NP}}$ $=$ 0.72$^{+0.15}_{-0.10}$, indicating that, if anything, \bedit{systems in our sample of polluted white dwarfs} have fewer giant planets. \bedit{However,  it is not clear that our “non-polluted” sample is in fact devoid of polluted white dwarfs (see \S\ref{sec:caveats}). Further work, such as that described in \S\ref{newsample}, is therefore required to place confident constraints on the ratio of Jupiter frequencies in polluted and non-polluted systems. }

\bedit{Equation~\ref{eq:power_law} can also be used} to estimate the sensitivity of a sample to differences in the mean metallicities. Here we define sensitivity as the fraction of samples where we would expect to observe a significant difference in the mean metallicities of polluted and non-polluted systems. To calculate the sensitivity for a given $f_{\textrm{P}}$, we use a Monte Carlo simulation. We randomly generate 50,000 samples containing 18 polluted and 41 non-polluted systems (representative of our own sample) and assume a constant value of $f_{\textrm{NP}} = 0.07$ (consistent with the observed mean metallicity of such systems in our sample and assuming $M_{\star}= 1.2$ $M_{\odot}$, consistent with our estimated mean progenitor mass). We then find the fraction of samples that have mean polluted and non-polluted metallicities that are significantly (1$\sigma$) different. For instance, if all polluted white dwarfs have massive planets ($f_{\textrm{P}}=1.0$), then we expect to observe a significant metallicity dependence in nearly all samples. We show the sensitivity as a function of $f_{\textrm{P}}$ in  Figure~\ref{jup_freq_samples}. We also show the probability of finding a metallicity difference ($\mu_{\textrm{P}} - \mu_{\textrm{NP}}$) greater than the one we observe ($-0.04$) as a function of the Jupiter frequency in polluted systems in the bottom panel of Figure~\ref{jup_freq}.

\bedit{There are several caveats to this calculation (in addition to those pertaining to our sample selection, as described in \S\ref{sec:caveats}). We note that }Equation~\ref{eq:power_law} is derived using a catalogue of planets with masses greater than $0.5$ M$_{\textrm{J}}$ found via the radial velocity technique. Because radial velocity detections are biased toward more massive planets at smaller orbital separations, Equation~\ref{eq:power_law} is limited to predicting the frequency of close-in giant planets. However, we expect that any close-in planets around a white dwarf progenitor star will be engulfed as the star transitions off the main sequence. We are therefore interested in the population of giant planets at large orbital separations. In order to compare the Jupiter frequencies around polluted and non-polluted white dwarfs, we make the assumption that Equation~\ref{eq:power_law} can be extrapolated to larger separations. 

Additionally, this relation is calculated excluding any known stars in binary systems, and is therefore not representative of our sample of white dwarfs in binaries. However, because our sample is dominated by wide binaries, we expect minimal interaction between the white dwarf and its binary companion. Furthermore, while close-in binaries can suppress the planet frequency \citep{kra2016,moe2021}, 
wide binaries are not believed to affect the prevalence of close-in (P$<$300 days) planets \citep{dea2015}. We therefore do not expect the binarity of our target systems to have a significant effect on their Jupiter frequency.

\section{Discussion}
\label{sec:discussion}
\subsection{White Dwarf Pollution Mechanisms}
We do not identify a significant metallicity \bedit{or accretion rate} dependence on white dwarf pollution. This study therefore does not find any evidence that massive planets are a major contributor to white dwarf pollution. \bedit{We note that this result is limited by several experimental caveats (see \S\ref{sec:caveats}) that could potentially be addressed in future work (see \S\ref{newsample}).}
\bedit{However, our preliminary findings} are consistent with theoretical work finding that less massive planets are more efficient inward scatterers \citep[e.g.,][]{bon2011, fre2014}, meaning that Jovian planets are not needed to explain observed white dwarf accretion rates. Our results are also consistent with \citet{blo2022}'s findings that white dwarf accretion rates do not experience a strong decrease over several Gyrs. Because systems with massive ($>100 M_{\oplus}$) planets are most dynamically active in the first Gyr of cooling time \citep{mal2020}, giant planets are insufficient to account for the observed accretion rates of older ($\gtrsim1$ Gyr) white dwarfs.

Our work also probes the importance of MMRs between rocky bodies and Jupiter-mass planets. This mechanism, proposed by \citet{deb2012}, would perturb the orbits of the rocky material, causing them to be tidally disrupted by the host star and eventually accreted onto the surface of the white dwarf. Because this requires the presence of a gas giant, our results indicate that this mechanism is not the primary source of white dwarf pollution in our sample of white dwarfs in wide binaries.

\bedit{We note that, even if future studies with more robust samples (e.g., as described in \S\ref{newsample}) corroborate our findings, we will not be able to} distinguish between the importance of non-planetary effects and scattering events caused by less massive planets, because smaller planets show a much weaker metallicity effect.

\subsection{Prospects for Comparing Stellar and Planetary Compositions}
Beyond studying the origin of white dwarf pollution, our sample (and existing observations) could also be used to probe the refractory compositions of stars and planetary systems. Within the Solar System, rocky planetary bodies and chondritic meteorites tend to have a refractory content consistent with the Sun's \citep[e.g.,][]{and1982}. However, it is uncertain whether this trend holds for other planetary systems. Using the abundances of a binary companion as a proxy for the primordial abundances of a polluted white dwarf, a white dwarf's composition can be compared to the composition of the rocky bodies accreted onto its surface. \citet{bon2021} demonstrated this approach for the binary system containing the K-dwarf G200-40 and the polluted white dwarf WD 1425$+$540. The authors found that the abundances of the K-dwarf and accreted planetary material are consistent within observational errors. This comparison could be expanded to a larger sample of polluted white dwarf binary systems, including the sample used in this work, to investigate the relationship between host star abundances and their rocky companions.

\subsection{Observational Follow-Up}
While measuring primordial metallicity serves as an indirect measurement of the prevalence of Jupiter analogues in white dwarf systems,  direct detections of planets would allow for more definitive comparisons between polluted and non-polluted white dwarf systems. In the past few years, a handful of planets and candidates have been detected around white dwarfs \citep{siggurdson2003, luhman2011, gansicke2019, vanderburg2020, blackman2021}, making it possible that directly probing the impact of planets on white dwarf pollution will be feasible in the future. 

{\it Gaia} catalogues will provide improved constraints on gas giant occurrence rates in white dwarfs systems. It is predicted that {\it Gaia} will find tens of thousands of exoplanets beyond 1 \bedit{au} with its high precision astrometry \citep{cas2008, per2014, ran2018}. In fact, one exoplanet candidate had already been found with {\it Gaia} EDR3 \citep{gai2022}, but was recently retracted\footnote{https://www.cosmos.esa.int/web/gaia/dr3-known-issues\#FalsePositive}. Additionally, \citet{san2022} \bedit{predicts 8 $\pm$ 2 astrometric detections of exoplanets orbiting white dwarfs.} These exoplanets are expected to have masses 0.03 \bedit{-} 13 $M_{\textrm{J}}$ and semi-major axes of 1.6 \bedit{-} 3.91 \bedit{au}. \bedit{If we assume 25\% of white dwarfs are polluted \citep{koe2014}, we would expect $\sim$2 of these newly discovered systems to be polluted, an insufficient number for a statistical survey of polluted and non-polluted systems. }

We will also be able to detect distant planets with the {\it Nancy Grace Roman Space Telescope}. {\it Roman} will use gravitational lensing to search for exoplanets that are smaller and more distant than those found by traditional transiting and radial velocity techniques. In total, Roman is expected to find $\sim$1400 planets with gravitational microlensing \citep{pen2019}. Because microlensing does not require any measurable light from the host star to detect planets, it will be sensitive to stars orbiting faint hosts, including white dwarfs \citep{blackman2021}.
This will provide an opportunity to search for both rocky and giant planets around white dwarfs, providing further insight into the dynamics of these evolved systems. \bedit{However, because we will not be able to measure the spectra of the faint microlensing host stars, we will not be able to determine whether they are polluted or not.}

The {\it James Webb Space Telescope} (JWST) can also help characterize white dwarf planetary systems. The Mid-Infrared Instrument (MIRI) Medium Resolution Spectrograph is capable of detecting cold, unresolved exoplanets via infrared excesses, and cold, resolved exoplanets via direct imaging. \citet{lim2022} predicted that JWST/MIRI will be able to detect Jupiter-sized planets for 34 of the closest white dwarfs. \bedit{It is} sensitive to planets at all orbital separations, and would provide insight into the demographics of white dwarf planetary systems. Several programs to search for planets around nearby white dwarfs have already been approved \citep{mul2021, meow, mead}\bedit{, providing constraints on the presence of planets \citep{pou2023} and identifying planet candidates \citep{mul2024}. Because it may be possible to discover a relatively large sample of planets around white dwarfs using JWST, this approach is likely the best way to statistically determine whether giant planets drive pollution. }

\section{Conclusions} \label{sec:conclusion}
25 - 50$\%$ of white dwarfs are metal polluted, evidence that material containing heavy elements recently accreted onto their surface. Polluted white dwarfs provide insight into the composition of rocky bodies and the physical processes governing the system. However, while many accretion mechanisms have been proposed, the dominant underlying mechanisms remain uncertain. 

To investigate the mechanisms driving this pollution, we observe a sample of polluted and non-polluted white dwarf binary systems. Using the companion stars' metallicities as a proxy for the white dwarfs' primordial metallicities and the giant planet population orbiting these stars, we compare the metallicity distributions \bedit{and inferred accretion rates} of polluted and non-polluted systems. We find no significant difference between the distributions, indicating that Jupiter analogs likely do not play a significant role in driving white dwarf pollution. While our results are limited by our small sample size \bedit{and non-uniform sample}, they are consistent with theoretical studies predicting that the amount of material scattered inward is either independent of, or weakly inversely related to, the mass of the scattering planet \citep[e.g.,][]{bon2011, fre2014}. This suggests that, if planet-driven scattering contributes to white dwarf pollution, it may be frequently caused by Earth- or Neptune-sized planets.

\section*{Acknowledgements}

This material is based upon work supported by the National Science Foundation Graduate Research Fellowship under Grant No. 1745302. This research has made use of NASA's Astrophysics Data System Bibliographic Services, and the SIMBAD database, operated at CDS, Strasbourg, France. Additionally, this work has made use of data from the European Space Agency (ESA) mission
{\it Gaia} (\url{https://www.cosmos.esa.int/gaia}), processed by the {\it Gaia}
Data Processing and Analysis Consortium (DPAC,
\url{https://www.cosmos.esa.int/web/gaia/dpac/consortium}). Funding for the DPAC
has been provided by national institutions, in particular the institutions
participating in the {\it Gaia} Multilateral Agreement. This publication also makes use of data products from the Two Micron All Sky Survey (2MASS) and the Wide-field Infrared Survey Explorer (WISE). 2MASS is a joint project of the University of Massachusetts and the Infrared Processing and Analysis Center/California Institute of Technology, funded by the National Aeronautics and Space Administration and the National Science Foundation. WISE is a joint project of the University of California, Los Angeles, and the Jet Propulsion Laboratory/California Institute of Technology, funded by the National Aeronautics and Space Administration.

\section*{Data Availability}
Derived parameters are provided in Tables~\ref{tab:SPClist}, ~\ref{tab:targetlist}, and ~\ref{tab:accretion}. Spectra are available upon request to the authors.
 



\bibliographystyle{mnras}
\bibliography{bib} 


\clearpage
\appendix
\onecolumn
\begin{table}
\section{\bedit{Inferred Accretion Rates}}
\begingroup 
\begin{center}
\centering
\caption{\bedit{List of all white dwarfs with available Ca abundances. We include the companion and white dwarf TIC IDs, Ca abundances [Ca/H(e)], sinking timescales $\tau$, and total inferred accretion rates $\dot{M}$. We additionally distinguish between accretion rates that are upper limits (Y) and that are estimated values (N). We provide references for Ca abundances obtained from literature sources. \bedittwo{Abundances with no listed reference were estimated in this work, as described in \S\ref{sec:accretion}.}}} \label{tab:accretion}
\centering
\begin{tabular}{ccccccc}
\hline
\hline
 Companion & White Dwarf  & [Ca/H(e)]  & $\tau$ & $\dot{M}$ & Upper& Reference\\
 ID &  ID &   &  &  & Limit?& \\
 & & (dex) &  (s) & (g s$^{-1}$) & & \\
\hline
21860383 & 21860382 & $-$6.2  & $6.14 	\times 10^{5}$ & $7.14 	\times 10^{8}$& Y & \\ 
 162690651 & 1201039482 & $-$9.0  & $1.22 	\times 10^{12}$ & $1.27 	\times 10^{8}$ & N& 1 \\ 
 458455132 & 166668903 & $-$6.5  &$8.25 	\times 10^{4}$ & $2.62 	\times 10^{8}$ &  Y& \\ 
 125838647 & 427771898 & $-$10.5 &$9.46 	\times 10^{11}$ & $3.45 	\times 10^{6}$ &  Y & \\ 
 284628177 & 284628173 & $-$9.1  &$2.69 	\times 10^{13}$ & $5.61 	\times 10^{7}$&  N & \\ 
 445863106 & 1001309484 & $-$9.9 &$1.97 	\times 10^{14}$ & $3.31 	\times 10^{6}$ &  N & 2 \\ 
 219857965 & 1271265755 & $-$8.5  &$4.13 	\times 10^{11}$ & $2.23 	\times 10^{8}$ &  Y&  \\ 
 305267214 & 2000305088 & $-$10.6  & $2.76 	\times 10^{13}$ & $3.25 	\times 10^{6}$& N & 2 \\ 
 427863040 & 427863042 & $-$7.3 & $6.24 	\times 10^{9}$ & $1.29 	\times 10^{9}$  & Y& \\ 
 9726029 & 9726026 & $-$6.3  &$1.39 	\times 10^{5}$ & $3.33 	\times 10^{8}$ &  Y& \\ 
 439869953 & 439869954 & $-$8.7 & $8.44 	\times 10^{10}$ & $9.66 	\times 10^{7}$& Y  &  \\ 
 422890692 & 630273894 & $-$7.2  &$1.34 	\times 10^{12}$ & $5.20 	\times 10^{9}$&  N & \\ 
 35703955 & 630323435 & $-$11.1  &$7.21 	\times 10^{15}$ & $1.19 	\times 10^{5}$&  N & 2 \\ 
 270371546 & 630406417 & $-$9.9  &$1.27 	\times 10^{15}$ & $8.87 	\times 10^{6}$ &  N& 2 \\ 
 336892483 & 610972154 & $-$5.7 &$1.54 	\times 10^{14}$ & $1.42 	\times 10^{11}$ &  N & 3 \\ 
 36726370 & 36726368 & $-$9.5 &$2.69 	\times 10^{11}$ & $2.26 	\times 10^{7}$ &  Y& \\ 
 352817358 & 352817378 & $-$10.4 &$6.23 	\times 10^{12}$ & $3.73 	\times 10^{6}$ &  Y & \\ 
 376455973 & 611128637 & $-$10.2 & $1.5 	\times 10^{15}$ & $1.35 	\times 10^{6}$  & N& 2 \\ 
 268127217 & 611261822 & $-$8.5  &$7.96 	\times 10^{12}$ & $3.0 	\times 10^{8}$ &  N& 4 \\ 
 172703279 & 172703276 & $-$7.9  &$2.76 	\times 10^{10}$ & $3.13 	\times 10^{8}$&  Y & \\ 
 129904995 & 129905006 & $-$9.8  &$6.38 	\times 10^{11}$ & $1.43 	\times 10^{7}$ &  Y& \\ 
 47350711 & 47353019 & $-$6.2  &$2.14 	\times 10^{7}$ & $1.08 	\times 10^{9}$ &  Y& \\ 
 1732730 & 841989747 & $-$11.0  &$3.95 	\times 10^{13}$ & $1.22 	\times 10^{6}$ &  N& \\ 
 191145234 & 191145238 & $-$7.0  &$3.2 	\times 10^{9}$ & $1.59 	\times 10^{9}$ &  Y& \\ 
 347389665 & 347389664 & $-$5.6  &$6.88 	\times 10^{8}$ & $2.9 	\times 10^{10}$ &  Y& \\ 
 239209564 & 903441740 & $-$11.1  &$8.67 	\times 10^{13}$ & $7.94 	\times 10^{5}$ &  N& 2 \\ 
 318811655 & 1204769581 & $-$6.2  &$9.19 	\times 10^{4}$ & $4.99 	\times 10^{8}$ &  Y&  \\ 
 149692037 & 149692034 & $-$6.8  &$1.58 	\times 10^{5}$ & $9.15 	\times 10^{7}$&  Y & \\ 
 277255071 & 277255078 & $-$10.8  &$9.33 	\times 10^{13}$ & $1.93 	\times 10^{6}$&  Y & \\ 
 23395123 & 23395129 & $-$8.9 &$6.12 	\times 10^{11}$ & $7.91 	\times 10^{7}$ &  Y & \\ 
 241194061 & 241194059 & $-$4.9&$3.18 	\times 10^{6}$ & $1.36 	\times 10^{10}$  &  Y & \\ 
 17301096 & 900244294 & $-$8.8 &$4.63 	\times 10^{13}$ & $5.39 	\times 10^{7}$ &  N & 2 \\ 
 280310048 & 471011543 & $-$11.8&$4.83 	\times 10^{14}$ & $5.52 	\times 10^{4}$  &  N & 5 \\ 
 \hline
 \multicolumn{7}{l}{%
  \begin{minipage}{12cm}%
   \bedit{References: (1) \citet{blo2022}; (2) \citet{cou2019}; (3) \citet{mar2023}; (4) \citet{car2023}}; (5) \citet{farihi2022} 
  \end{minipage}%
}\\
\end{tabular}
\end{center}
\endgroup 
\end{table}
\section{Comparison to Planet Search Stars and Milky Way Field Stars}
\label{sec:field_comparison}
Because our goal is to make inferences about planet populations based on the average metallicities measured in samples of white dwarfs, we compare the metallicity distribution of our targets with the metallicity distribution of planet search host stars. In particular, we compare our targets to stars observed in the California Legacy Survey \citep[CLS,][]{lee2021}\bedit{. Overall}, the mean metallicity of the full CLS sample ($\mu_{\textrm{CLS}}=-0.01\pm0.01$, $\sigma_{\textrm{CLS}}=0.28\pm0.01$) is $0.16 \pm 0.03$ dex higher than the mean metallicity of our full sample ($\mu=-0.17\pm0.03$, $\sigma=0.24\pm0.02$). \bedit{In addition, the median of the CLS sample (0.03 dex) is 0.21 dex above the median of our sample ($-$0.18 dex).} Another large radial velocity planet search program, the Retired A-star survey \citep{ghe2018}, also shows a significantly higher mean metallicity ($\mu_{\textrm{RAS}}=-0.07\pm0.01$, $\sigma_{\textrm{RAS}}=0.201\pm0.009$) than our sample. 

 \begin{table}
\centering
\footnotesize
\label{tab:obslog}
\centering
\caption{\bedit{Metallicity distributions of planet host stars. For each catalog, we provide the stellar types and sample size, in addition to the median M$_{\rm[m/H]}$, mean $\mu_{\rm[m/H]}$, and standard deviation $\sigma_{\rm[m/H]}$ of each metallicity distribution. Our sample has a median metallicity of $-$0.18 dex, a mean metallicity of $-$0.17 $\pm$ 0.03 dex, and a standard deviation of 0.24 $\pm$ 0.02 dex.  }} \label{tab:catalog_comparisons}
\begin{tabular}{lrrr}
\hline
 \multirow{4}{*}{}  & CLS & RAS & GC  \\
 \cline{2-4}
 Stellar Types & FGKM & Subgiants & FG \\
Sample Size & 719 & 245 & 16,394 \\ 
M$_{\rm[m/H]}$ & 0.03 & $-$0.06 & $-$0.14 \\
$\mu_{\rm[m/H]}$ & $-$0.01 $\pm$ 0.01  & $-$0.07 $\pm$ 0.01  & $-$0.167 $\pm$ 0.002  \\
$\sigma_{\rm[m/H]}$ & 0.28 $\pm$ 0.01 &0.201 $\pm$ 0.009 & 0.237 $\pm$ 0.001 \\
 \hline
\end{tabular}
\end{table}
 
By contrast, our low metallicity values are not significantly different from the metallicities of a broader sample of stars in the solar neighborhood from the Geneva-Copenhagen (GC) survey \citep{nor2004, holm2007, holm2009}. This large spectroscopic survey measured the \bedit{rotations, ages, kinematics, and Galactic orbits of 16,682 F and G dwarfs. A subsample of 16,394 stars in the GC catalog have metallicity measurements.} The mean metallicity of the GC sample \bedit{is $-0.167\pm0.002$ dex, consistent with our results. However, we note that the sample has a low-metallicity tail and that the median of the distribution is $-$0.14 dex.} The mean metallicity of the GC sample is \bedit{$-0.16\pm0.01$} dex lower than the mean metallicity of the CLS sample and \bedit{$-0.10\pm0.01$} dex lower than the Retired A-star sample, indicating that these planet search catalogs may be biased toward higher metallicities. \bedit{We summarize the three catalogs used for comparison (CLS, RAS, and GC) in Table~\ref{tab:catalog_comparisons}.}

We consider several possible reasons why the planet search samples, and CLS in particular, have preferentially higher metallicity than both our white dwarf sample and the GC sample, but are ultimately unable to identify a satisfactory explanation. We first investigate whether this discrepancy can be partly attributed to our selection of binary systems. CLS draws most of its targets from the Keck Planet Search \citep[KPS,][]{cumm2008}, the Eta-Earth Survey (EES) \citep{how2010}, and the 25 pc northern hemisphere sample \citep{hir2021}, and excludes known binary systems. As discussed in \S\ref{sec:separationrelation}, we do not expect wide binaries (with separations $>$250 \bedit{au}) to display a metallicity offset \citep{elb2019}. To further probe the relation between binarity and metallicity, we use the GC sample of 16,682 F and G dwarf stars. We cross-list the survey with the catalog of {\it Gaia} \bedit{EDR3} binaries. We find $\sim$1,000 binary systems with a mean metallicity of $-0.1162\pm0.002$ dex, and $\sim$15,000 non-binary systems with a mean metallicity of $-0.14857\pm0.00001$ dex. Both samples have a median metallicity of approximately $-0.11$ dex. As GC non-binary systems tend to have a slightly lower metallicity, the exclusion of binary systems is likely not biasing CLS toward higher metallicities. 

Additionally, while the CLS catalog does not explicitly apply color or magnitude restrictions, KPS and the 25 pc sample both require $B - V > 0.55$, and EES requires $V < 11$. Using the GC stars that meet these criteria, we find $\sim$200 binary systems with a mean metallicity of $-0.0966\pm0.0007$ and $\sim$1,700 non-binary systems with a mean metallicity of $-0.1421\pm0.0002$. The combined mean metallicity is $-0.1351\pm0.0001$, $-0.0112\pm0.0001$ higher than the mean metallicity in the absence of color and magnitude restrictions. This difference is not large enough to account for the discrepancy between the CLS and GC mean metallicities.

We also consider whether the age of the observed systems may be contributing to the metallicity offset between our sample and CLS. Because we require that one of the binary stars has had sufficient time to evolve off the main sequence, it's possible our sample is biased toward older stars. However, as discussed in \S\ref{sec:relations}, there does not appear to be a significant dependence on age for local white dwarfs in binary systems.

Regardless of the origin of the difference in mean metallicity between the planet search sample and our white dwarf sample, the significant offset makes it challenging to interpret our results in the context of planet occurrence. \bedit{However, we can still use the CLS results to predict an expected metallicity difference between systems with and without Jupiters (see \S\ref{sec:prediction_comparison}).}


\bsp	
\label{lastpage}
\end{document}